\documentclass{aa}  
\usepackage{graphicx}
\usepackage{lipsum}
\usepackage{lscape}             
\usepackage{placeins}           
\usepackage{enumitem}
\usepackage{booktabs}
\usepackage{longtable}
\usepackage{graphicx}
\usepackage[varg]{txfonts}
\usepackage[breaklinks=true]{hyperref}
\usepackage{siunitx}
\usepackage{float}
\usepackage{orcidlink}

\usepackage{mathtools}
\usepackage{textgreek}
\usepackage{lastpage}
\usepackage{supertabular}
\usepackage{multirow}
\usepackage{natbib}
\usepackage{tikz}
\hypersetup{
    colorlinks=true,
    linkcolor=blue,
    filecolor=magenta,      
    urlcolor=cyan,
    citecolor=blue
    }
\makeatletter
\renewcommand*\aa@pageof{, page \thepage{} of \pageref*{LastPage}}
\makeatother

\begin{document}

   \title{A New Perspective on Galactic Evolution: Studying the Outskirts
of the Abell S1063 Galaxy Cluster}

   \subtitle{}

   \author{L. Pecoraro 
          \inst{1,} \inst{2}\orcidlink{0009-0007-9726-2646}\and
          A. Mercurio\inst{1,}\inst{2,}\inst{3}\orcidlink{0000-0001-9261-7849}\and
          M. Annunziatella\inst{4}\orcidlink{0000-0002-8053-8040}\and
          M. D'Addona\inst{2}\orcidlink{0000-0003-3445-0483}\and
          R. Ragusa\inst{2}\orcidlink{0009-0000-6680-523X}\and
          G. Angora\inst{2}\orcidlink{0000-0002-0316-6562}\and
          M. Girardi\inst{6,7}\orcidlink{0000-0003-1861-1865}\and
          G. Granata\inst{8}\orcidlink{0000-0002-9512-3788}\and
          C. Grillo\inst{4, 5}\orcidlink{0000-0002-5926-7143}\and
          L. Limatola\inst{2}\orcidlink{0000-0002-1896-8605}\and
          P. Rosati\inst{9, 10}\orcidlink{0000-0002-6813-0632}\and
          P. Bergamini\inst{10}\orcidlink{0000-0003-1383-9414}\and
          G. Caminha\inst{11, 12}\orcidlink{0000-0001-6052-3274}\and
          F. Getman\inst{1}\orcidlink{0000-0003-1550-0182}\and
          A. Grado\inst{13, 14}\orcidlink{0000-0002-0501-8256}\and
          M. Meneghetti\inst{10}\orcidlink{0000-0003-1225-7084}\and
          E. Vanzella\inst{10}\orcidlink{0000-0002-5057-135X}
          }

   \institute{Università di Salerno, Dipartimento di Fisica “E.R. Caianiello”, Via Giovanni Paolo II 132, I-84084 Fisciano (SA), Italy
         \and
             INAF – Osservatorio Astronomico di Capodimonte, Via Moiariello 16, I-80131 Napoli, Italy
         \and
             INFN – Gruppo Collegato di Salerno - Sezione di Napoli, Dipartimento di Fisica “E.R. Caianiello”, Università di Salerno, via Giovanni Paolo II, 132 - I-84084 Fisciano (SA), Italy 
         \and
             INAF – IASF Milano, via A. Corti 12, I-20133 Milano, Italy
         \and
             Dipartimento di Fisica, Università degli Studi di Milano, via Celoria 16, 20133 Milano, Italy
         \and
              Dipartimento di Fisica dell’Università degli Studi di Trieste – Sezione di Astronomia, Via Tiepolo 11, I-34143 Trieste, Italy 7
         \and 
              INAF – Osservatorio Astronomico di Trieste, Via Tiepolo 11, I34143 Trieste, Italy
         \and
              Institute of Cosmology and Gravitation, University of Portsmouth, Burnaby Rd, Portsmouth PO1 3FX, UK
         \and
              Dipartimento di Fisica e Scienze della Terra, Università degli Studi di Ferrara, Via Saragat 1, 44122 Ferrara, Italy
         \and
             INAF – OAS, Osservatorio di Astrofisica e Scienza dello Spazio di Bologna, Via Gobetti 93/3, I-40129 Bologna, Italy
         \and
             Max-Planck-Institut für Extraterrestrische Physik, Giessenbachstr. 1, 85748 Garching, Germany
         \and
             TUM School of Natural Sciences, Technical University of Munich, Garching 85748, Germany
         \and
             Dipartimento di Fisica e Geologia, Università degli Studi di Perugia, Via Alessandro Pascoli, s.n.c., 06123 Perugia, Italy
         \and    
             INFN Sezione di Perugia, Via Alessandro Pascoli, s.n.c., 06123 Perugia, Italy
             }


   \date{Received ...; accepted xxx, 2025}


  \abstract
   {\textit{Context}. Galaxy physical properties are influenced by their environments, but the processes responsible for mass and environmental quenching and structural transformations remain debated. Galaxy clusters are ideal laboratories for investigating galaxy formation and evolution, offering a full range of galaxy properties and environments. Observations of large-scale structures, particularly filaments in cluster outskirts ($r \sim$5$r_{200}$), are currently constrained to the low-redshift Universe. To explore galaxy evolution at intermediate redshifts, deep photometric data, ideally combined with spectroscopic redshifts, are essential. \\
   \textit{Aims.} Abell S1063 cluster ($z=$ 0.346) is observed within the Galaxy Assembly as a function of the Mass and Environment program with the VLT Survey Telescope (VST-GAME) combined VISTA Public Survey program Galaxy Cluster At Vircam. We investigate galaxy evolution across a wide range of stellar masses and environments. We release a multiwavelength photometric catalog with photometric redshifts for 64394 sources over a 1$\times$1 deg$^2$ field of view. The analysis of overdensity regions provides insights for future studies on galaxy properties in cluster outskirts. \\    
   \textit{Methods.} The dataset is obtained through deep ($r<$24.65 mag) and wide optical ($u$, $g$, $r$, $i$, VST) and near-infrared ($Y$, $J$, $Ks$, VISTA) observations.  The photometric catalog includes all detected sources, excluding nearby or overlapping objects, saturated stars, and image artifacts. The multiwavelength catalog enabled the photometric redshifts measurements for the cluster membership. The density field measurement allowed comparison of galaxy properties, colors and masses, across different cluster environments. \\
   \textit{Results.} We detect a very dense structure near the cluster center and thanks to such a large field of view, we find another dense region to the north-west, in the opposite direction to the cluster elongation. Filaments connecting the regions are also visible. }
   \maketitle

\section{Introduction}
Galaxy clusters represent invaluable astrophysical testbeds for studying the key physical processes driving galaxy formation and evolution. 
Hosting diverse populations, from massive core galaxies to faint dwarfs, they allow for investigating how properties are shaped by both mass and environment. Clusters at intermediate redshifts ($z\sim$0.4) are particularly relevant for observing the impact of the hierarchical formation of cosmic structures on the evolution of galaxy properties. Models suggest that over the past 5 Gyr, galaxies have undergone rapid and significant evolution \citep[e.g.][]{Butcher_e_Olmer1978, Butcher_e_Olmer1984}. At redshifts $z\sim$0.2-0.4, clusters predominantly host spiral galaxies with active star formation, which may have been progressively replaced by more evolved systems in the local Universe \citep{Dressler97, Treu03}. This evolutionary trend suggests that gas-rich star-forming spirals were accreted into clusters at $z\sim$0.5-1 and subsequently transformed into the passive galaxies observed in present-day clusters. Gas depletion or suppression driven by one or more cluster-related mechanisms is responsible for structural transformations in galaxies. 
\\Observational evidence indicates that, at least since $z<$ 1, both the current properties and the past evolution of galaxies are strongly influenced by their environment \citep[environmental-quenching, e.g.,][]{Blanton05, Tanaka05, Peng10, Lemaux19}. Environmental mechanisms such as tidal stripping \citep[e.g.][]{Zwicky_1951,Gnedin_2003}, harassment \citep[e.g.][]{Spitzer51, Moore96, Moore99}, group and cluster collisions, and starvation \citep[e.g.][]{Larson80, Boselli06} play crucial roles in shaping galaxy evolution. At high velocities, galaxies can undergo ram pressure stripping \citep[see][]{Sheen17, Foltz18}, a process that effectively removes a substantial fraction of their cold gas. 
Moreover, the fraction of quenched galaxies strongly depends on the galaxy stellar mass \citep[mass-quenching, e.g.,][]{Van_der_Burg20}. The star formation histories and morphologies of massive galaxies are primarily determined by their assembly through mergers, with significant Active Galactic Nuclei and Supernovae feedback \citep[e.g.][]{Gòme03, Tanaka_2004, Haines_2006, Fritz09}. On the other hand, dwarf galaxies are more strongly influenced by environmental factors. Passive dwarf ellipticals, for example, are typically found as satellites within massive halos, whether in clusters, groups, or around large galaxies \citep[][]{Haines_2006, Boselli_2014, Fritz14, Roberts19}. Despite decades of work \citep[e.g.][]{Dressler80, Kauffmann04, Balogh04, Postman_2005, Boselli06, Smith_2005, Mercurio10, Moresco10, Kovač13, Annunziatella14, Just15, Jaffè15, Annunziatella16, Rhee_2017, Oemler_2017, Owers_2019, Rhee20}, the importance of these processes in the transformation of galaxies as a function of mass and environment remains debated. Recently, in an attempt to investigate these processes, the outskirts of galaxy clusters have emerged as critical testbeds of cosmology and cluster astrophysics \citep[]{Estrada23, Ragusa25}. The characterization of these regions is essential to understanding the physics of the intracluster medium (ICM) beyond the central core. Galaxies moving through the hot, ionized ICM undergo various quenching processes that inhibit or suppress subsequent star formation. Variations in ICM properties from the cluster core to the outskirts \citep[][]{Nagai11, Ichikawa_2013, Lau_2015, Biffi18, Mirakhor21} lead to different degrees of interaction of galaxies in these different environments, and the efficiency of the physical processes may also be different. Notably, infalling galaxies in outer regions may undergo accelerated or complete gas removal before reaching the core \citep[][]{Zabludoff98, Mihos04, Fujita04}. Studying these regions at intermediate and high redshifts is particularly important, as the mechanisms governing galaxy evolution differ significantly from those observed in the local Universe \citep[][]{Van_der_Burg20}. At these redshifts ($\sim$ 0.5-1), clusters mainly accrete blue spiral galaxies and gas-rich star-formers through filaments \citep[][]{Colberg99, Ebeling_2004, Castignani_2021} and less in the other directions.Galaxies accreted along high-density filaments experience more rapid quenching than those infalling from other directions \citep[][]{Martinez16, Salerno19, Salerno20}. So, to fully understand the evolution of galaxy properties, it is essential to observe and analyze a large sample of galaxies across the entire cluster environment, from the core to the outskirts and infall regions and trace the rapid evolution of galaxy populations during these epochs \citep[][]{Poggianti_2006, Desai_2007}.
\\In this work, we present a detailed study of the galaxy cluster Abell S1063 \citep[hereafter AS1063,][]{Abell89} also cataloged as RXJ2248.7-4431 as it was detected in the ROSAT All-Sky Survey \citep[][]{DeGrandi99, Guzzo99}, out to its external region. AS1063 is a very massive cluster at $z=$0.3457 \citep[][]{Mercurio21} with a total mass value of (2.9$\pm$0.3)$\times$10$^{15}$ M$\odot$ \citep[][]{Sartoris20}, a high X-ray luminosity ($L_X \approx$ 8$\times$10$^{45}$ erg s$^{-1}$) and high X-ray temperature \citep[$T_X\approx$ 13 KeV,][]{Gómez_2012}. Previous works have investigated AS1063, including the weak lensing analysis \citep{Gruen13}, the examination of the Kormendy relation \citep{Tortorelli18}, the detection of a Sunyaev-Zel'dovich signal with high significance in the Planck data \citep[][]{PlanckCollaboration11, Plagge_2010}, and the study of the dynamics of the different components of the galaxy \citep[][]{Ferrami_2023}. Additionally, \citet{Xie20} identified a giant radio halo with a size of  $\simeq$1.2 Mpc and an integrated spectral index that steepens in the range from 1.5 to 3.0 GHz.
\\AS1063 was also part of the Hubble Space Telescope (HST) treasury program Cluster Lensing And Supernova Survey with HST, \citep[CLASH,][]{Postman12} and the Hubble Frontier Fields initiative \citep[][]{Lotz17}. Additionally, the cluster AS1063 was observed with Visible Multi Object Spectrograph (VIMOS) as part of the ESO Large Programme 186.A-0798 "Dark Matter Mass Distributions of Hubble Treasury Clusters and the Foundations of $\Lambda$CDM Structure Formation Models" \citep[hereafter CLASH$-$VLT, PI: P. Rosati,][]{Rosati14}, a large-scale spectroscopic survey of 13 CLASH clusters visible from ESO-Paranal. The CLASH-VLT observations were further complemented in the cluster core with the Multi Unit Spectroscopic Explorer (MUSE) \citep[][]{Bacon17, Mercurio21, Ferrami_2023}.
\\The analysis of an unprecedentedly sample of 1234 spectroscopic member galaxies out to 1.7$r_{200}$\footnote{$r_{200}$ is the radius of the sphere which the mean density is 200 times the critical mass density of the universe at the cluster redshift.} ($r_{200}=$2.63 Mpc) in \citet{Mercurio21}, pointed out that AS1063 is far from being dynamically relaxed. Evidence includes a bimodal velocity distribution, a NE-SW elongation and evidence of substructures suggesting a SE-NW merger. According to \citet{Gómez_2012}, based on Gemini Multi-Object Spectrographs (GMOS) spectroscopic data for 51 members and Chandra X-ray data, AS1063 has experienced a recent merger event near the plane of the sky along the NE$-$SW direction. This elongation, evident in both X-ray emission and dark matter (DM) distributions, show distinct morphologies: while the DM is approximately centered on the brightest cluster galaxy (BCG), the hot gas is offset toward the northeast and appears to be more rounded. Thus, AS1063 is an excellent target for investigating environmental impacts on galaxy evolution.
\\In this work, we exploit the optical and near-infrared data from Very Large Telescope (VST) and Visible and Infrared Survey Telescope for Astronomy (VISTA), covering for the first time an extensive region out of $r=$3.8$r_{200}$ (or $R \approx$ 10.0 Mpc) from the cluster center. This dataset allows us to analyze galaxy populations across a wide range of environmental densities, spanning from the dense core to the diffuse outskirts. We present the final photometric catalog of the AS1063 cluster.
\\The paper is structured as follows: in Section~\ref{sec:2} we describe observations and data reduction. In Section~\ref{sec:3} we outline source extraction, masking, and catalog construction (detail in Appendix~\ref{app:A}). In Section~\ref{sec:4} we present the photometric redshift estimation (see also Appendix~\ref{app:B}) and cluster member identification. In Section~\ref{sec:5} we describe the red sequence and the density field, while in Section~\ref{sec:6} we discusses and summarizes our results.
\begin{figure*}[t!] 
    \centering 
    \includegraphics[width=0.6\textwidth]{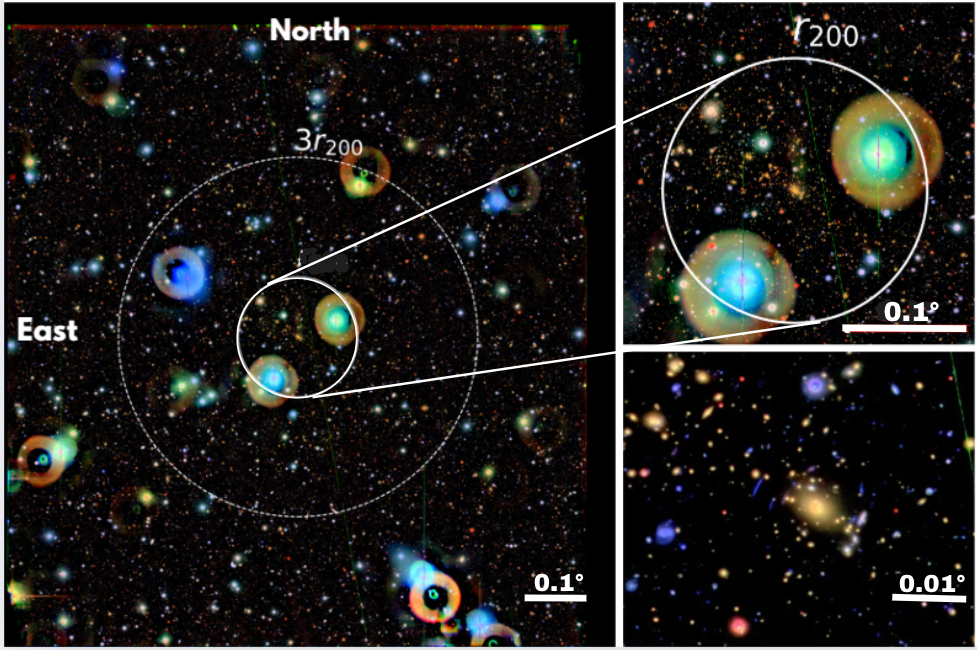}
    \caption{\small Left panel: VST color composite image of AS1063 made using $g$, $r$, and $i$ band as color channels blue, green, and red, respectively (East on the left and North on the top). The solid white circle indicates the $r_{200}$ and the dashed one indicates the 3$r_{200}$. Right top panel: A zoomed view of the cluster in within $r_{200}\simeq$ 2.63 Mpc. Right bottom panel: Zooming into the central part of the cluster. The red lines represent the scale.} 
    \label{fig:2248} 
\end{figure*}
\\Throughout the paper, we adopt a standard $\Lambda$CDM cosmology with $\Omega_m = $0.3, $\Omega_\Lambda =$ 0.7, and $H_0 =$ 70\, $\mathrm{km\,s^{-1}\,Mpc^{-1}}$. At the redshift of the cluster, $z=$ 0.3457, 1 arcmin corresponds to 0.294 Mpc. The magnitudes are given in the AB system, and all distance measurements are comoving distances. The cluster center coordinates are \text{RA} = 342.1830357 and \text{Dec} = $-$44.5307675. We use $r_{200} =$ 2.63 $\pm$ 0.09 Mpc as in \citet{Sartoris20}.

\section{Observations and data reduction}
\label{sec:2}

\subsection{VST-VISTA dataset}
\label{sec:2.1}
AS1063 (Fig.~\ref{fig:2248}) was observed in the $u$, $g$, $r$, $i$ optical bands with VST as a part of the Galaxy Assembly as a function of Mass and Environment with VST survey (P.I. A. Mercurio, hereafter VST-GAME) and in the Near Infrared $Y$, $J$, $Ks$ with VISTA as a part of the VISTA Public Survey program Galaxy Cluster At Vircam (P.I.: M. Nonino, hereafter G-CAV). 
\\To investigate galaxy evolution down to a stellar mass of $10^9$ $M_\odot$, VST-GAME performs a unique wide-field coverage (20$\times$ 20 Mpc$^2$) of 12 massive galaxy clusters, in a redshift range between 0.2 and 0.6 ($z$ median $\sim$ 0.4). The VST-GAME survey began during the ESO period P99, using 300h of the Italian INAF Guaranteed Time Observations with OmegaCAM. G-CAV observed 20 galaxy clusters, covering $\sim$30 square degrees. Both surveys are completed by spectroscopic data observed as a part of the CLASH-VLT survey with a field of view of $\sim$25$\times$25 \text{$arcmin^2$}.
\\ This deep multi-band and high-quality data set enables the selection of cluster members using photometric redshifts, the characterization of the environment through local galaxy density, the potential identification of structures like filaments, and the study of galaxy properties as a function of galaxy density in the cluster.

\subsection{Data reduction}
\label{sec:2.2}
Data reduction follows the methodology of \citet{Estrada23}. The optical data obtained with the VST were reduced using the VST-Tube pipeline \citep{Grado2012}. The standard procedure of image reduction, alignment, and co-addition was carried out by applying over-scan correction, bias subtraction, and flat-field calibration. A weight map was produced for each exposure, accounting for the reliability of individual pixels. The reduced CCD frames were astrometrically calibrated using Scamp \citep[][]{Bertin06} with Gaia DR2 as the reference system. Individual frames were then resampled and co-added into final stacks using \texttt{Swarp} \citep{Bertin02} to ensure global astrometric consistency. To guarantee a uniform photometric response, was applied an illumination correction map derived by comparing the point-source catalog with Pan-STARRS \citep{Chambers16}. The G-CAV survey is a second-generation ESO VISTA Public Survey. The near-infrared images used in this work were reduced, calibrated, and stacked by the G-CAV survey team using the standard VISTA data reduction pipeline. We retrieved fully processed images from the ESO Science Portal\footnote{\url{http://archive.eso.org/scienceportal/home?data_collection=GCAV&publ_date=2020-12-07h}} and the ESO Science ArchiveFacility\footnote{\url{http://archive.eso.org/programmatic/##TAP}}
Data reduction was performed independently for each photometric band. In Table~\ref{tab:table1}, we report the total exposure times for the entire dataset and the FWHMs of the PSF for each band used for the photometric analysis.

\section{Photometry and catalog construction}
\label{sec:3}
In this paper, we analyze the spectro-photometric dataset of AS1063, which offers a unique photometric coverage across a wide wavelength range [0.32-2.37] $\mu$m, thanks to observations from VST and VISTA. Building a complete catalog is a crucial starting point to classify cluster members from the dense core to the outskirts, with a construction procedure specifically optimized for this purpose.
\\In this section, we describe the procedure for catalog extraction, star-galaxy separation, and source masking. We provide an overview of the photometric analysis and highlight the differences respect to the analysis of \citet{Estrada23} necessitated by the specific characteristics of this dataset.
\begin{table}[h]
\centering
\begin{tabular}{cccccc} 
\toprule 
 Band & Exp. time & FWHM &  Completeness  & Zero-point \\
     & [ks]       & [arcsec] &  90$\%$[mag] & [mag] &  \\
        \midrule 
        $u$ & 63.18 & 1.40 &  25.35 & 30.00\\ 
        $g$ & 53.04 & 0.86 &  25.15 & 30.00\\ 
        $r$ & 77.87 & 0.66 & 24.65 & 30.00\\ 
        $i$ & 47.10 & 0.83 & 23.30 & 30.00\\ 
        $Y$ & 28.80 & 0.77 &  23.70 & 29.60\\ 
        $J$ & 21.12 & 0.75 &  23.45 & 30.21\\ 
        $Ks$ & 16.80 & 0.83 & 22.85 & 30.36 \\ 
        \bottomrule 
    \end{tabular}
    \caption{\small VST and VISTA observational dataset.}
    \label{tab:table1}
\end{table}

\subsection{Source detection}
\label{sec:3.1}
\begin{figure*}[t!]
    \centering
    \includegraphics[width=1\textwidth]{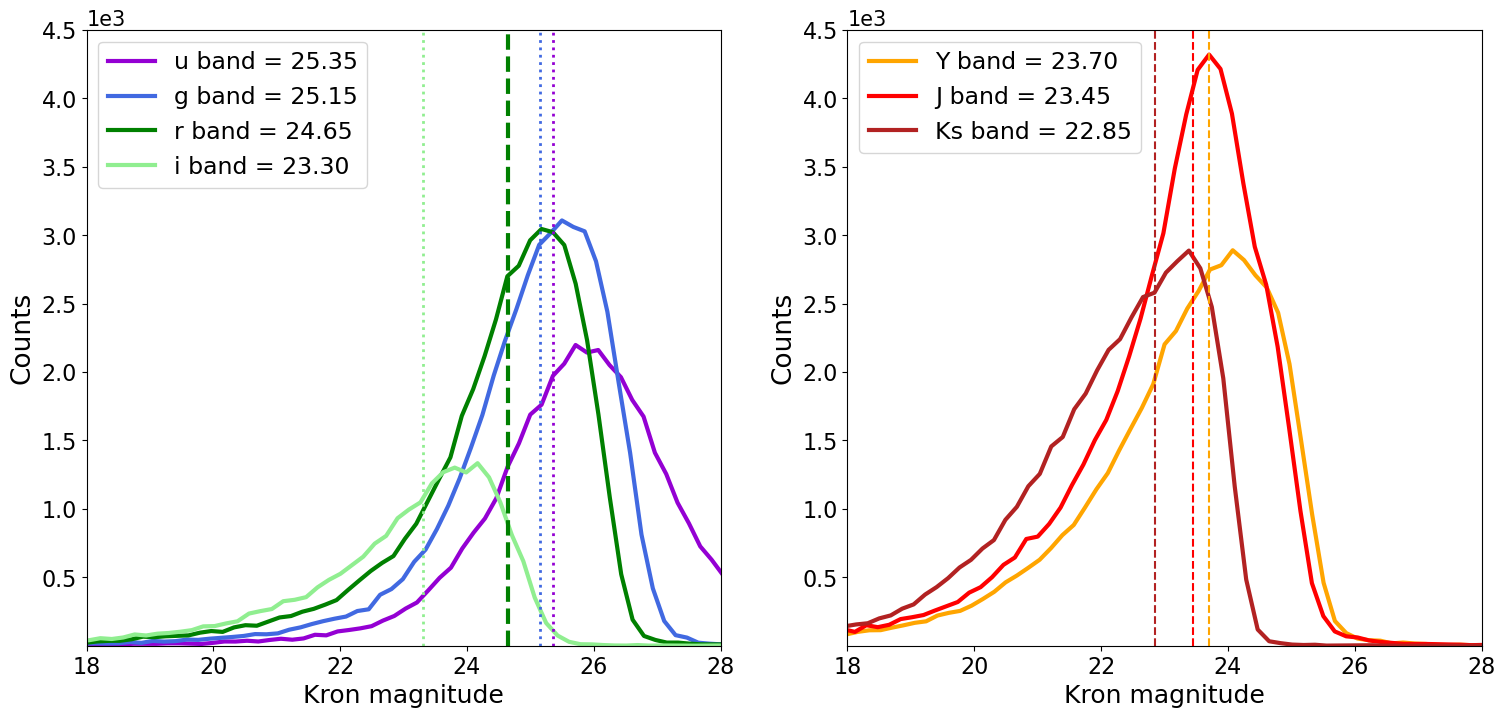}  
    \caption{\small Left panel: magnitude distribution of sources extracted with SExtractor for $u$, $g$, $r$, $i$ optical bands. Right panel: magnitude distribution for $Y$, $J$, $Ks$ near-infrared objects. The vertical lines represent the completeness limits for each band (see Table~\ref{tab:table1}).}
    \label{fig:compl}
\end{figure*}

To detect both bright galaxies and a substantial number of faint sources, we optimized the source extraction process, paying particular attention to avoiding nearby or overlapping sources in crowded fields and ensuring that large, bright galaxies were identified as single objects. 
\\The VST and VISTA images were processed using SExtractor \citep[][]{Bertin96} and PSFEx \citep[][]{Bertin11}, which accurately models the point spread function (PSF). SExtractor uses the PSF models generated by PSFEx to improve PSF-corrected model-fitting photometry for all sources within the image. 
Source extraction can be performed either in single–image mode, where both source detection and photometric measurements are carried out on the same image, or in dual–image mode, in which the detection is performed on a reference band while photometry is measured on the band of interest. In this work, we adopt single–image extraction for the $g$, $r$, $i$, $Y$, $J$, and $Ks$ bands, as these images have sufficiently high signal–to–noise ratios to ensure reliable detection and photometry within the same band. The exception is the $u$-band, for which the catalog is extracted in dual–image mode using the $r$-band as the detection image. This choice is motivated by the lower signal–to–noise ratio of the $u$-band data, which can limit reliable source detection. Using the deeper $r$-band image for detection improves the completeness of the catalog while ensuring consistent source positions and apertures across bands, enabling the identification of faint source.
\\ We adopted the same SExtractor configuration for VST images, described in Table 2 of \citet{Estrada23}, for a similar dataset. The only difference is the \texttt{BACK\_SIZE} value, where we assume 128 instead of 64. The \texttt{BACK\_SIZE} keyword is crucial for accurately estimating the background in astronomical images and, consequently, obtaining precise flux and size measurements of the sources. If the value is too small, it may not adequately capture background variations, while a value too large may encompass nearby sources, distorting the calculation. We derived aperture photometry within ten different apertures in arcsec: 1.5, 2, 3, 4, 8, 16, 30, 45, 3*FWHM, 8*FWHM. The FWHM values used are listed in Table~\ref{tab:table1}.
\\ We compute the completeness magnitude limit of the extracted catalogs as the magnitude limit below which we start to lose fainter galaxies. We adopt the approach of \citet{Garilli99}, which was also applied in \citet{Mercurio15}. In Table~\ref{tab:table1}, the completeness magnitude limit at 90$\%$ for each band is indicated, while Fig.~\ref{fig:compl} shows the magnitude distribution of all sources detected with SExtractor. We find 227964 sources for the $r$ and $u$ bands, 230481 for the g band, 107684 for the $i$ band, 263288 for the $Y$ band, 332107 for the $J$ band, 243837 for the $Ks$ band. Due to the larger field of view of VISTA, the number of detected objects is larger compared to those observed in the optical bands, except for the $r-$band, which presents a longer exposure time. In the left panel of Fig.~\ref{fig:compl}, we present the VST optical bands together with their completeness limit, while the right panel illustrates the VISTA bands. Throughout this work, all magnitudes refer to \texttt{MAG\_AUTO} unless otherwise stated.
\subsection{Star-galaxy separation}
\label{sub:3.2}
To ensure the reliability and accuracy of the photometric analysis, special attention is given to remove saturated stars and spurious detections from each catalog, thus improving the accuracy of the star–galaxy separation. During source extraction and data processing, errors can occur due to failed deblending, incomplete isophotal measurements, and/or memory overflow issues. Star–galaxy separation is carried out through an iterative procedure, using four parameters to distinguish stars from galaxies. A flag is assigned to each object in the catalog based on four key parameters, which are used to distinguish between stars and galaxies. The \texttt{CLASS\_STAR} (or stellarity index) parameter, a continuous star–galaxy classifier ranging from 0 to 1, where values close to 1 correspond to point-like sources (stars) and values close to 0 to extended sources (galaxies). With this classification, sources are identified as stars if they have $r\geq$15.75 mag (i.e., brighter than the saturation limit) and a stellarity index $\geq$0.5, or if $r\leq$15.75 mag and the stellarity index $\geq$0.99. However, \texttt{CLASS\_STAR} alone is insufficient for classifying faint objects. To further distinguish between stars and galaxies, we use \texttt{FLUX\_RADIUS\_50}, the radius of the circular aperture enclosing 50\% of the total flux of the object, which provides a measure of the source's size. Galaxies, being more extended, typically exhibit larger half-flux radius values compared to point-like stars. The SPREAD\_MODEL parameter quantifies the dispersion of an object's light relative to the PSF model obtained with PSFEx. This parameter is particularly useful for extending classification to fainter magnitudes. Stars, being more concentrated, exhibit a \texttt{SPREAD\_MODEL} value close to 0, whereas more diffuse galaxies show a larger spread. Finally, we use the MU\_MAX parameter, which represents the maximum surface brightness of an object per unit of area. This parameter helps distinguish between saturated and unsaturated stars, confirming classifications made using the previous parameters. Saturated stars are identified below the limit of 15.75 mag in the $r$ band. The stars selected in the previous steps present a clear trend with the Kron magnitude \citep{Kron76}, following a line with higher \texttt{MU\_MAX} values compared to the more diffuse galaxies.
\\ Star-galaxy separation procedure was performed in all optical bands. In this way, each source in each catalog is labeled with a flag of 0 for galaxies, 1 for stars, and values greater than 1 for saturated sources and spurious objects identified by the different parameters. Sources with a flag value of 0 in the $r$ band are confidently identified as galaxies, while those with flag values greater than zero are classified as stars. Furthermore, a more detailed analysis was performed using the software SAOImage DS9 to examine objects flagged as 0 in the $r$-band. A random subsample of approximately 70 objects identified as galaxies, scattered throughout the frame, is examined to ensure that they are not saturated stars or deblended objects. This examination confirms that objects with flag\_r$=$0 are indeed galaxies and validates that the $r$ band serves as a reliable reference band.

\subsection{Masking of spurious regions}
\label{sub:3.3}
\begin{figure}
    \centering 
    \includegraphics[width=0.49\textwidth]{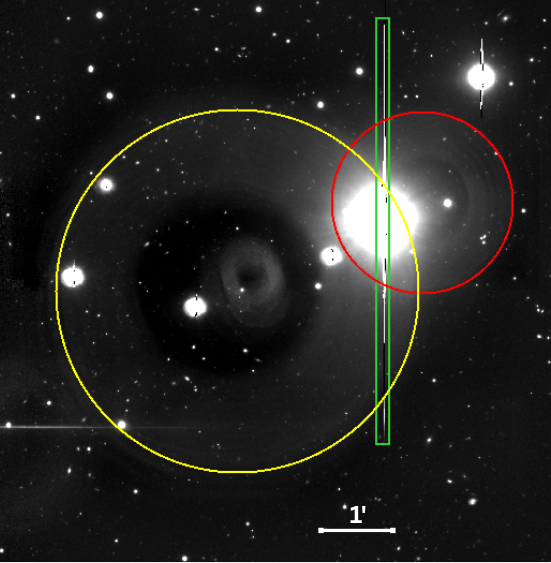} 
    \caption{\small Examples of artifact by a saturated star in the $r$-band image, are illustrated: the vertical diffraction spike is in green, the halo produced by the star in red, the ghost in yellow. A white scale bar provides a reference of 1 arcmin.} 
    \label{fig:spike} 
\end{figure}

The presence of bright stars in images can introduce artifacts in the field of view as an increase of surface brightness in some regions.
As a result, nearby sources may have flux measurements that are affected by these irregular patterns. For this reason, it is crucial to accurately identify these major impurities as spikes, halos, and ghosts. In Fig.~\ref{fig:spike}, spikes are marked by narrow vertical lines of light extending from the star (green rectangle), halos are circular regions of increased brightness around the star (red circle), and ghosts consist of a central area with lower surface brightness, surrounded by a brighter outer corona (yellow circle). To mask halos and ghosts, we utilize the position and magnitude of the stars responsible for these artifacts. To accurately determine the magnitude and the centers of these sources, we compare our catalog with Gaia Data Release 3 \citep[][]{Gaia18}, trying to obtain automatic circles around the halo and rectangles for the spike, based on the Gaia magnitude. We verify that the halos, ghosts, and spikes are generated by saturated stars in optical bands, while in the near-infrared band, also not saturated stars can produce ghosts. With a general procedure, we automatically generate circular masks covering halos and ghosts, as detailed in Appendix~\ref{app:A}.
\\ We observe a correlation between the Gaia G magnitude of the brightest stars and the radii of the halos they produce: the radius increases by 0.5 pixels for each magnitude step, towards fainter stars. In Tables~\ref{tab:mask_VST} and~\ref{tab:mask_VISTA}, we report the Gaia magnitude cutoffs and the corresponding halo radii for each band. The saturation levels of each band are 99.0 for the $u$-band, 13.0 mag for the $g$-band, 13.9 mag for the $r$-band, and 12.9 mag for the $i$-band, 11.6 mag for the $Y$-band, 13.2 mag for the $J$-band, and 13.2 mag for $Ks$-band.
\\ To verify which stars produce ghosts, we consider the Gaia magnitude and the limiting magnitude of the stars in the different bands, finding values of 10.3 mag for the $u$-band, 11.0 mag for the $g$-band, 11.5 mag for the $r$-band, 11.4 mag for the $i$-band, 11.5 for the $Y$-band, 12.9 mag for the $J$-band and 9.1 mag for the $Ks$-band. Subsequently, we find a relation that links the distance from the center of the field and the center of the ghost and the distance from the center of the field and the center of the star for each band and we note that also the size of the ghosts changes as reported in Table~\ref{tab:mask2} in Appendix~\ref{app:A}.
\\The spikes are identified through manual detection, as they are asymmetric and their size does not appear to correlate with the magnitude of the stars that generate them. Spikes associated with stars completely embedded into halos within halos and/or ghosts, which do not involve other sources, are not masked. 
\\After the masking procedure on the multi-band sample, we find for the reference band $r$ that 29262 sources (12.8$\%$) are included the masks (\texttt{FLAG\_MASK}$=$1 if they are in halos, 2 if ghosts and 3 if spikes), while 198695 sources (87.2$\%$) are outside the masks (\texttt{FLAG\_MASK}$=$0). In Table~\ref{tab:mask_source} of Appendix~\ref{app:A}, we report all the values of the masked sources for each band.

\subsection{Multiwavelength catalog} 
\label{sec:3.4}
In this work, we present a high-quality multi-band catalog, generated by matching all sources from the $r$-band catalog within the completeness limit of r$\geq$24.65 mag. The seven catalogs, including masking and star-galaxy classification flags, were combined into a complete optical-NIR photometric catalog containing 64394 objects, including both stars and galaxies and with the Kron magnitude in all bands. After matching with the spectroscopic catalog, we note that 231 objects have the wrong stellarity index. Therefore, we estimate that the contamination of our catalog is about 0.36$\%$. 
\\The zero-points are set to 30.0 mag for the VST telescope while the zero-point of VISTA are calibrated using F105W and F125W from HST for $Y$ e $J$ bands and F200W from JWST for the $Ks$ band (see Appendix~\ref{app:B} for JWST data reduction). The average photometric error is less of 0.1 mag, demonstrating a good level of measurement precision. In Table~\ref{tab:table1}, we report the values of the zero points after correction.
\\The matching process was performed using STILTS \citep[][]{Taylor06}, with a maximum separation of 0.5 arcsec. Table~\ref{tab:cat} presents an extract of the final catalog. The \texttt{GAME\_ID} uniquely identifies each source, along with its barycentric coordinates \texttt{RA} and \texttt{Dec} in degrees. The parameters \texttt{A} and \texttt{B} represent the semi-major and semi-minor axes of the ellipse describing the object’s shape, while \texttt{THETA} denotes the position angle in degrees. The \texttt{FLUX\_RADIUS\_50} and \texttt{KRON\_RADIUS} values correspond to the half-light radius and Kron radius in pixels, both derived from the $r$-band. In this extraction, we include only the magnitudes \texttt{MAG\_APER\_15}, \texttt{MAG\_APER\_30}, and \texttt{MAG\_APER\_40}, which represent aperture magnitudes with diameters of 1.5, 3, and 4 arcsec, respectively. \texttt{MAG\_AUTO} refers to the magnitude, which provides an estimate of the "total magnitude" by integrating pixel values within an adaptively scaled aperture, within \texttt{KRON\_RADIUS}, \texttt{MAG\_MODEL} is the model magnitude derived from the sum of the spheroidal and disk components, and \texttt{MAG\_PSF} is the magnitude calculated using the PSF. The magnitudes of the various bands may be weaker than their corresponding completeness limits since the individual catalogs were not clipped off before the match. The \texttt{SI} column provides the stellarity index, obtained from SExtractor in the $r$-band. \texttt{FLAG\_BAND} contains the flag used in star-galaxy classification as described in Sect.~\ref{sub:3.2}: the zero value indicates galaxies, while values greater than zero denotes stars. \texttt{FLAG\_MASK} identifies objects located within stellar halos or spikes; as described in Sect.~\ref{sub:3.3}, a value of one points to an object within the mask, while a value of zero indicates an object outside of it. Finally, we provide the photometric redshift, the \texttt{$z_{\text{phot}}$} derived as described in the next section.
\\Upon acceptance of the article, the photometric catalog (an extract of which is shown in Table~\ref{tab:cat}) will be released as described.

\section{Photometric redshifts and cluster membership}
\label{sec:4}
In this section we obtain high-quality multiband photometry, which enables us to assign accurate photometric redshifts to catalog sources. We also describe the procedure for determining photometric redshifts and the criteria used to identify candidate members of AS1063.

\subsection{Photometric redshifts}
\label{sec:4.1}
From the multi-band catalog, we selected galaxies with $r$-band flag of zero and applied a magnitude cut of $r\leq$24.65 mag, resulting in a sample of 44675 galaxies. To ensure the reliability of these data, we calibrated our photometry using the available spectroscopic catalog of 1234 galaxies in AS1063 \citep[][]{Mercurio21}, obtained as part of the CLASH-VLT survey, using spectroscopic redshift $z_{\text{spec}}$.
\\ The photometric redshifts, $z_{\text{phot}}$, were estimated using the Random Forest algorithm \citep[][]{Ho98, Random01}, a machine-learning algorithm based on an ensemble of decision trees that combines multiple independent predictions to improve accuracy and reduce overfitting; the model was trained on a sample of 1508 spectroscopic galaxies and validated using 1006 spectroscopic objects that were not used during the training phase. Further details on the photometric sample selection and photometric redshift estimation are provided in Appendix~\ref{app:C}.
\\Figure~\ref{fig:crfz} shows the distribution of photometric redshifts compared to spectroscopic ones of the sample of 1006 objects. To evaluate the quality of photometric redshifts and identify objects with significant discrepancies, we employ the metric $\Delta z=$ ($z_{\text{phot}}-z_{\text{spec}}$)/($1+z_{\text{spec}}$) as in \citep[][]{Brescia19}. We defined as outliers all galaxies that have $|\Delta z|>$0.15, and the fraction of outliers, $\eta$, serves as a measure of the fit’s accuracy. We calculate three statistical parameters to evaluate the accuracy of the $z_{\text{phot}}$: the mean of $|\Delta z|$, representing the bias, which indicates the average deviation of the photometric redshifts from the spectroscopic values; the standard deviation of $|\Delta z|$, denoted as $\delta_{smooth}$; and the normalized median absolute deviation ($\sigma_{\text{NMAD}}$), calculated as 1.4826$\times$ median($|\Delta z|$), providing a robust measure of dispersion \citep[][]{Razim21}. The statistical results are: bias $=$ 0.006, $\delta_{smooth}$ $=$ 0.103, $\sigma_{\text{NMAD}}=$ 0.025 and $\eta$ $=$ 6.4$\%$.

\subsection{Selection of cluster members}
\label{sec:4.2}
Spectroscopic redshifts provide a critical constraints for optimizing photometric redshift ranges and accurately identifying galaxy members within the cluster. Figure~\ref{fig:range} presents the photometric redshifts distribution for objects with spectroscopic information. We first distinguish members (light orange) from objects with$z_{\text{spec}}$ (blue). Members are identified using the spectroscopic redshift range found in \citet{Mercurio21} of 0.333 $ \leq z_{\text{spec}} \leq $ 0.359 (solid red line) . Then, we select the photometric redshift range 0.304 $ \leq z_{\text{phot}} \leq $ 0.408. Completeness and purity are quantified for the selected photometric members. The completeness is defined as the fraction of spectroscopically confirmed cluster members that are also identified as photometric members, $\text{C} = N_{\mathrm{pm} \cap \mathrm{zm}} / N_{\mathrm{zm}}$, where $N_{\mathrm{pm} \cap \mathrm{zm}}$ is the number of galaxies identified as photometric members that are also spectroscopically confirmed members, and $N_{\mathrm{zm}}$ is the total number of spectroscopic members. The purity represents the fraction of photometric members that have a spectroscopic redshift, given by this $\text{P} = N_{\mathrm{pm} \cap \mathrm{zm}} / N_{\mathrm{pm} \cap \mathrm{z}}$, where $N_{\mathrm{pm} \cap \mathrm{z}}$ is the number of photometric members with available spectroscopic redshift.  We evaluate the completeness and purity of the selected photometric members. The chosen range, 0.304$ \leq z_{\text{phot}} \leq $0.408, provides an optimal balance between completeness and purity, for which we find the completeness of 91.3$\%$ and the purity of 69.0$\%$. 
\begin{figure} 
    \centering 
    \includegraphics[width=0.5\textwidth]{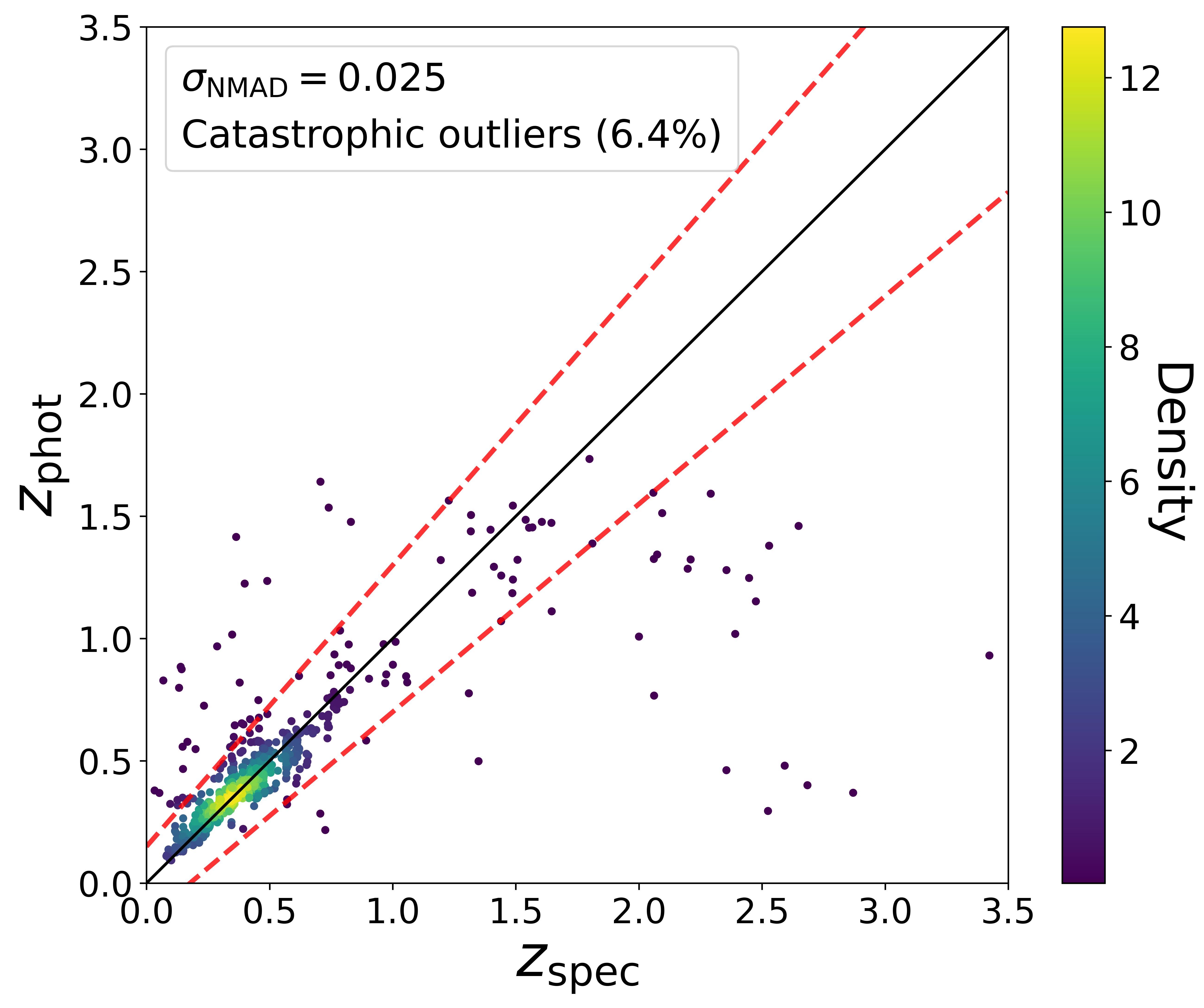} 
    \caption{\small Comparison between measured photometric $z_{\text{phot}}$ and spectroscopic $z_{\text{spec}}$ redshifts. The red dotted lines indicate the limit for the outliers ($|\Delta z| >$ 0.15). The fraction of outliers is $\eta = $6.4$\%$ while bias $=$ 0.006 and $\sigma_{\text{NMAD}} =$ 0.025. The bar shows the number density.} 
    \label{fig:crfz} 
\end{figure}
\begin{figure}
    \centering 
    \includegraphics[width=0.5\textwidth]{z_phot_range.jpeg}
    \caption{\small Photometric redshifts distribution for objects with available spectroscopic measurements. Spectroscopic cluster members are shown in light gold; all spectroscopic galaxies are in blue; $z_{cl} =$ 0.346 is the spectroscopic redshift of the cluster. The range of photometric membership, in green, is [0.304, 0.408].}  
    \label{fig:range} 
\end{figure}
\\ Although this selection implies the presence of contaminating interlopers, their impact on the analysis is expected to be limited. Our results rely on statistical properties of the galaxy population rather than individual objects, making them relatively insensitive to a moderate level of contamination. Moreover, the interlopers are not expected to introduce systematic biases, but contribute as a roughly uniform background component.
Increasing the photometric interval enhances completeness, by including more cluster members, but simultaneously increases the fraction of outliers, reducing purity. Across the entire VST field, among galaxies in the photometric catalog without measured spectroscopic redshifts, we identified 3693 galaxies that lie in the requested photometric range for membership. We refer to these galaxies as photometric members. Adding these to the 1234 spectroscopic member galaxies yields a total sample of 4927 member galaxies, also referred to as spectro-photometric members.

\section{Characterization of the environment}
\label{sec:5}
Galaxies exhibit a well-known morphological bimodality, broadly separating into elliptical and spiral types, and this distinction is also reflected in their colors. This bimodality is observed as a function of both stellar mass and local environment. The population of red elliptical galaxies dominates the densest regions of large-scale structures \citep[see Fig. 14 of ][]{Estrada23}, making them efficient tracers of the cosmic web. In the previous section, we defined a photometric redshift range used to identify cluster members. In this section, we compute the red sequence and analyze the density field of AS1063.

\subsection{Red sequence}
\label{sec:5.1}
\begin{figure}
    \centering
    \includegraphics[width=0.49\textwidth]{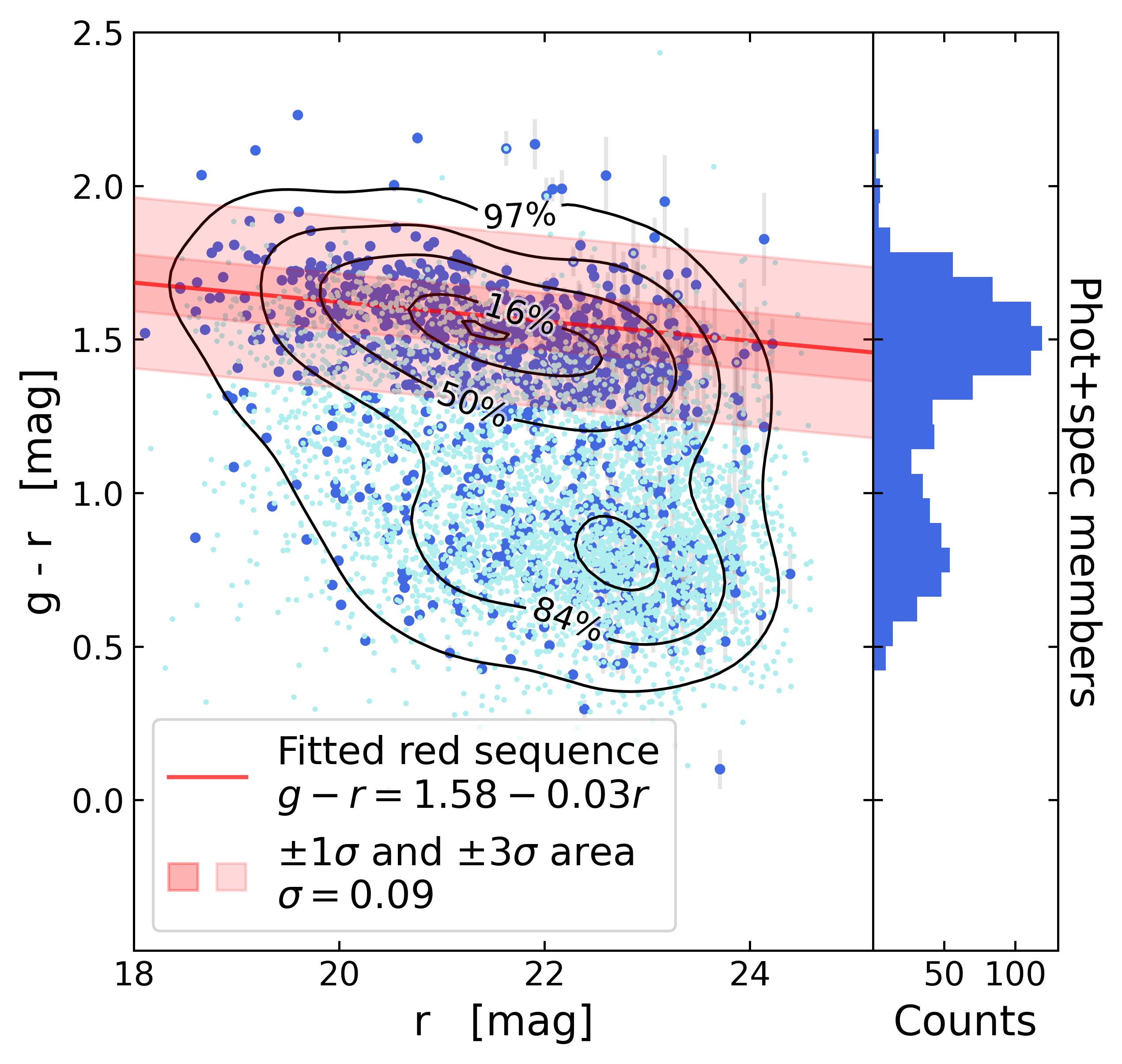}
    \caption{\small Color-magnitude diagram of the 1234 spectroscopically confirmed cluster members (dark blue) and of the 3693 photometric ones (light blue). The red line is the color–magnitude relation, with shaded regions indicating the $\pm1\sigma$ (light red) and $\pm3\sigma$ (dark red) limits. The right panel displays the color distribution of spectro–photometric members.}
    \label{fig:red}
\end{figure}

Here, we present the analysis of the color-magnitude diagram used to select red galaxies in the cluster field and to identify the overdensities. To derive the red sequence by using the $g-r$ color and $r$ magnitude, we employ the software from \citet{Cappellari13}. We fit a linear relationship to the spectroscopically confirmed member galaxies and the photometric samples. We then select red-sequence galaxies as those member both spectroscopic and photometric, that lie within the relation: $g-r =$ 1.58$-$0.03$r$, with a standard deviation of $\delta_{smooth}$ $=$ 0.09. Figure~\ref{fig:red} shows the resulting color-magnitude diagram with spectroscopic members in dark blue and photometric ones in light blue. The red solid line represents the best fit red-sequence, while the light and dark shaded regions indicate the $\pm1\sigma$ and $\pm3\sigma$ intervals. We count all photometric members within 3$\delta_{smooth}$ of the defined red sequence, identifying them as red sequence galaxies. At this confidence level, we find that there are 1333 red galaxies. Considering only red-sequence galaxies, we notice an improvement in the purity values of the photometric members selection. In fact, in the chosen redshift range, 0.304$ \leq z_{\text{phot}} \leq $0.408, we obtain a completeness of 100$\%$ and a purity of 72.2$\%$. This selection confirms that the choice of red galaxies exhibit more accurate photometric redshifts compared to the full sample, making them efficient tracers of dense structures \citep{Lubin20, Girardi20, Girardi22}.
\begin{table*}[t!]
\centering
\footnotesize 
\setlength{\tabcolsep}{5pt} 
\caption{First five lines of the final catalog in which are reported the values for $r$-band.}
\label{tab:cat}
\begin{tabular}{cccccccccc}
\toprule
\small GAME\_ID & RA & Dec & A  & B & THETA & KRON\_RADIUS & FLUX\_RADIUS\_50 & MAG\_AUTO & MAGERR\_AUTO \\
  & [deg] & [deg] & [pixel]  & [pixel] & [deg]  & [pixel] & [pixel] & [mag] & [mag] \\
\midrule
9455 & 342.3462 & $-$44.9789 & 19.078 & 12.367 & 88.380 & 3.50 & 14.522 & 12.1202 & 0.0004 \\
10309 & 341.4725 & $-$45.0006 & 8.7140 & 8.448 & $-$65.04 & 3.50 & 5.3810 & 13.7373 & 0.0001\\
10758 & 342.3384 & $-$44.9865 & 11.545 & 9.014 & 82.960 & 3.50 & 6.1920 & 13.6098 & 0.0004 \\
11826 & 342.0405 & $-$45.0070 & 3.1120 & 2.415 & 85.630 & 3.73 & 2.6260 & 22.2311 & 0.0105 \\
11866 & 342.2663 & $-$45.0068 & 4.5510 & 3.283 & $-$36.29 & 4.55 & 4.4010 & 22.2544 & 0.0182 \\
\bottomrule
\end{tabular}
\end{table*}
\vfill
\begin{table*}[t!]
\centering
\footnotesize 
\setlength{\tabcolsep}{6pt}
\label{tab:cat_mags}
\begin{tabular}{ccccccc} 
\toprule
\small MAG\_APER\_15 & MAG\_APERR\_15 & MAG\_APER\_30 & MAG\_APERR\_30 & MAG\_APER\_40 & MAG\_APERR\_40 & MAG\_PSF \\
{[mag]} & {[mag]} & {[mag]} & {[mag]} & {[mag]} & {[mag]} & {[mag]} \\
\midrule
15.3261 & 0.0001 & 13.8359 & 0.0001 & 13.3433 & 0.0001 & 99.0000 \\
15.2589 & 0.0001 & 14.1327 & 0.0001 & 13.9421 & 0.0001 & 99.0000 \\
15.3264 & 0.0001 & 14.1878 & 0.0001 & 13.9106 & 0.0001 & 99.0000 \\
22.7180 & 0.0060 & 22.3561 & 0.0083 & 22.2536 & 0.0100 & 22.2313 \\
23.2765 & 0.0099 & 22.6042 & 0.0104 & 22.4337 & 0.0118 & 22.9611 \\
\bottomrule
\end{tabular}
\end{table*}
\vfill
\begin{table*}[t!]
\centering
\footnotesize 
\setlength{\tabcolsep}{4pt}
\begin{tabular}{cccccccc}
\toprule
\small MAGERR\_PSF& MAG\_MODEL & MAGERR\_MODEL & CLASS\_STAR  & FLAG\_BAND & FLAG\_MASK & z\_phot  \\
 & [mag] & [mag] &  & & & \\
\midrule
99.0000 & 12.2125 & 0.0033 & 0.845 & 9.0 & 1 & 1.458 \\
99.0000 & 13.7373 & 0.0015 &  0.828 &  9.0 & 1 & 0.671\\
99.0000 & 13.6233 & 0.0039 & 0.845 &  9.0 & 1 & 1.159 \\
0.0049 & 22.2309 & 0.0478 &  0.891 & 2.0 & 0 & 1.011  \\
0.0091 & 22.2546 & 0.1564 &  0.027 &  0.0  & 0 & 0.690 \\
\bottomrule
\end{tabular}
\end{table*}
\begin{figure*}[t!] 
    \centering
    \includegraphics[width=0.81\textwidth]{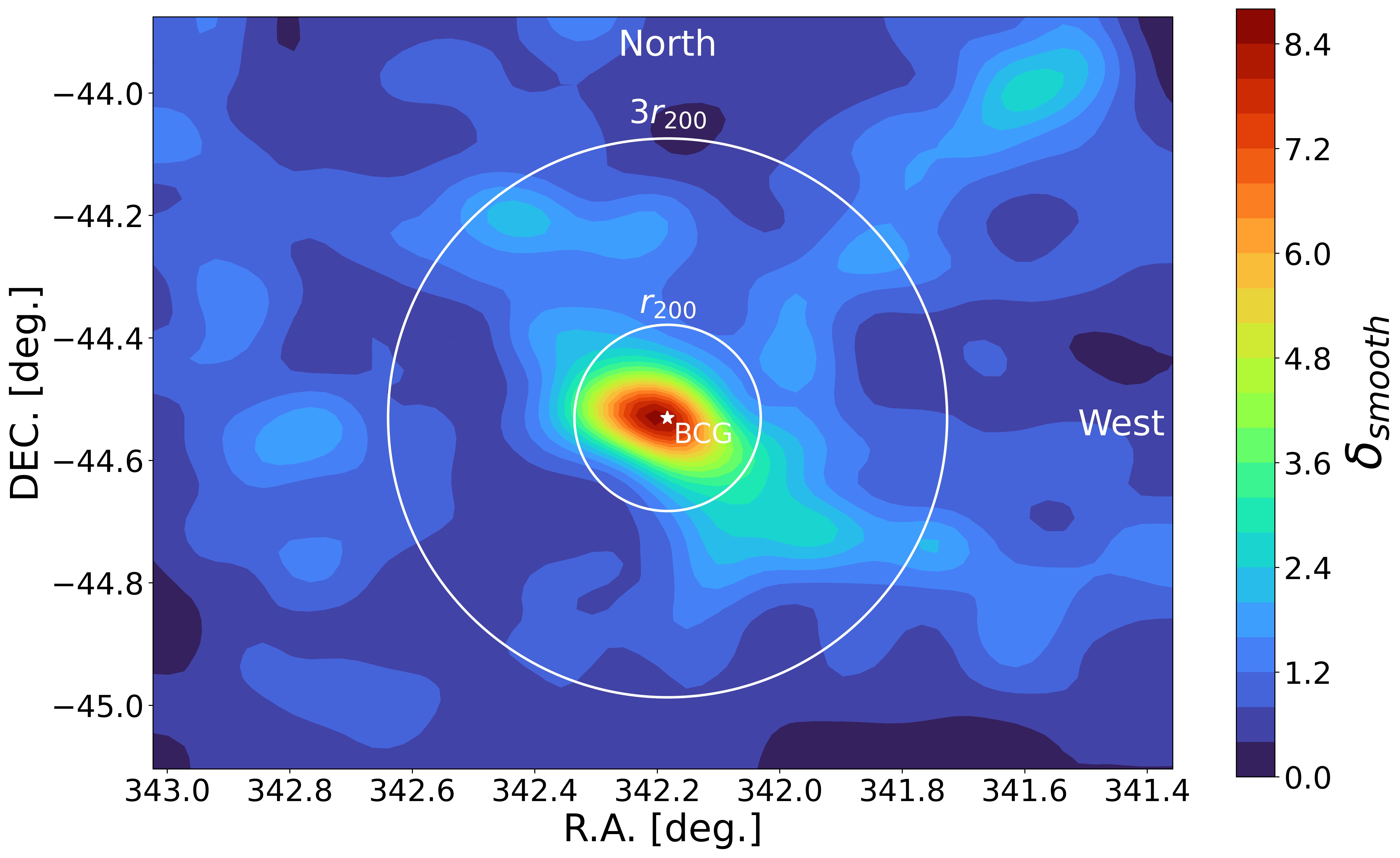}
    \includegraphics[width=0.81\textwidth]{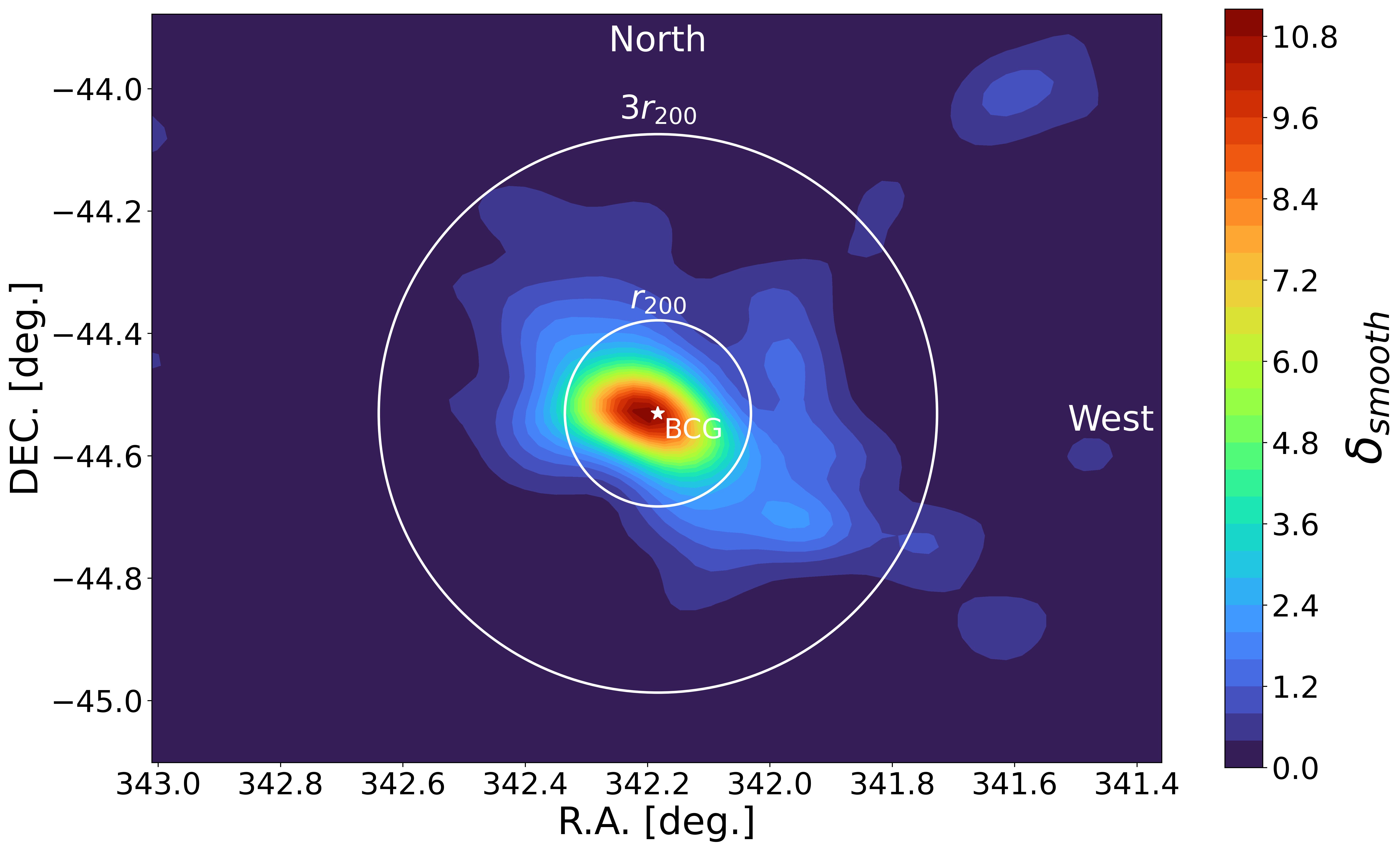}    
    \caption{\small \textbf{Top panel:} AS1063 density field computed using photometric and spectroscopic member galaxies. \textbf{Bottom panel:} density field obtained with the only red sequence galaxies within 3$\delta_{smooth}$. All galaxies are selected in the VST-GAME photometric redshift range 0.304$<z_{\text{phot}}<$0.408. The circles are centered around the BCG (white star): the inner circle indicates $r_{200}\simeq$2.63 Mpc and the outer one is $3r_{200}\simeq$7.89 Mpc.}
    \label{fig:density}
\end{figure*}

\subsection{Density Field}
\label{sec:5.2}
In this subsection, we compute the galaxy density field as a tracer of the different environments. Since the properties of galaxies vary as a function of the density of their surrounding environment, it is necessary to compute the galaxy density map of the field \citep[e.g.,][]{Gòme03, Kovac10, Annunziatella14, Granett15, Annunziatella16, Malavasi17, Gargiulo19, Ata21}. 
\\ We use both spectroscopic and photometric galaxies to define a rectangular grid with square cells of 1 arcmin per side, corresponding to a physical distance of approximately 300 kpc at the cluster redshift. Using a nearest-grid-point scheme, we count the number of galaxies in each cell ($n_i$) to compute the mean background density, $n_{mean}=$0.83 galaxies/arcmin$^2$. We then derive the $\delta_{smooth}$ quantity, obtained by smoothing the local density contrast $\delta_i=n_i/n_{mean}$ with a Gaussian kernel ($\sigma=1$ arcmin). Furthermore, we computed the density field using only red galaxies to be $n_{mean, rs}=$0.33 galaxies/arcmin$^2$ within 3$\delta_{smooth}$, following the same procedure. In general, the numerical values of the density field depend on the method used to calculate it, the cell size and the width of the Gaussian kernel which are chosen to find an optimal compromise between large-scale structures and small-scale numerical count fluctuations.
\\The top panel of Fig.~\ref{fig:density} shows the density field obtained using only photometric and spectroscopic member galaxies, while in the bottom panel, we use only red sequence galaxies. We note that a very dense clump outside $r_{200}$ and close to the cluster central region appears in both density maps, along with a secondary clump in the NW direction, in the opposite direction with respect to the NE-SW elongation of the cluster. As expected, the highest density values correspond to the central region of the cluster. Filaments connecting these substructures are also visible (top panel). We then checked the literature for known substructures outside 3$r_{200}$, with the aim of verifying whether the substructure we detect in the NW region had already been reported in previous works. We found no corresponding detections.

\subsection{Galaxy members in different environments}
\label{sec:5.3}
\begin{figure*}[t!]
    \centering
    \includegraphics[width=1\textwidth]{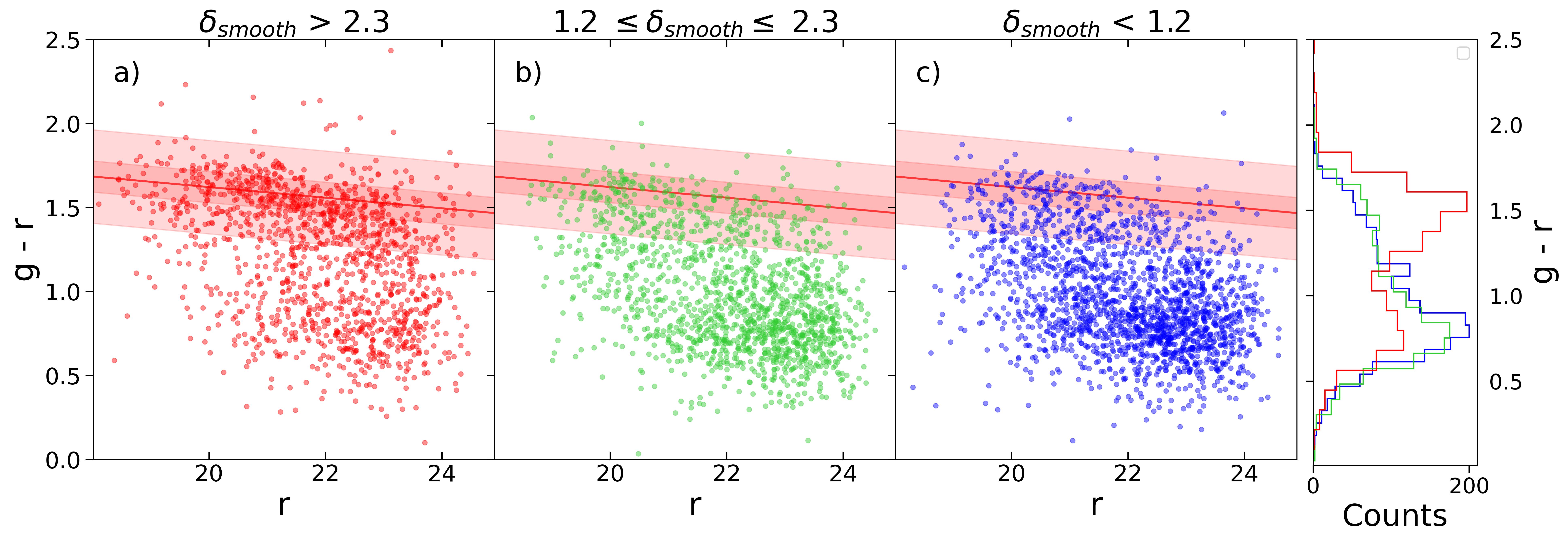}  
    \caption{\small Color-magnitude diagram: the spectro-photometric sample of galaxies subdivided according to the local environment, for each $\delta_{smooth}$ range chosen. Panel a): galaxies in environments dense than the field; Panel b): medium density with respect to the field; Panel c): galaxies in low density zones. The last panel on the right shows a histogram of the $g-r$ color, comparing the three galaxy samples. The number of red galaxies is 650, 329 and 354, respectively for the high, medium and low density environments. The red line represents the fit of red galaxies obtained with the sample of spectroscopic members, within 3$\delta_{smooth}$.}
    \label{fig:redsequence}
\end{figure*}

In order to investigate the local environment within the cluster, we divided the combined sample of photometric and spectroscopic member galaxies into three density ranges. This allows us to identify regions characterized by different levels of galaxy concentration in the kernel-smoothed density map (see Sect.~\ref{sec:5.2}).
We define low-density regions as $\delta_{smooth}<1.2$, intermediate-density regions as $1.2\leq\delta_{smooth}\leq2.3$, and high-density regions as $\delta_{smooth}>2.3$ galaxies/arcmin$^{2}$. These thresholds were adopted to separate galaxies located in diffuse environments from those associated with progressively denser regions of the cluster structure.
The results obtained show a distribution of 1882 galaxies (38.2$\%$) in the low density range, 1494 galaxies (30.3$\%$) in the medium density range, and 1551 galaxies (31.5$\%$) in the high density range.
The same procedure was applied to the subsample of 1333 red-sequence galaxies selected within $3\sigma$. In this case, we find 650 objects in high density regions, 329 in medium density regions, and 354 in low density regions.
\\ The different properties of galaxies in distinct local-density environments are analyzed by comparing the prominence of the red sequence, that is, the relative abundance of red galaxies within each of the three previously defined density ranges. Figure~\ref{fig:redsequence} shows the color-magnitude plot ($r$, $g-r$). We note that, in each sample, the galaxy population exhibits a bimodal color distribution, characterized by two distinct peaks. The relative importance of these two peaks varies significantly with the environmental conditions. Indeed, the peak of blue cloud decreases as the density increases. As shown in the distribution on the right side of Fig.~\ref{fig:redsequence}, the peak related to the redder colors corresponds to the galaxy population in the red sequence, while the blue peak is associated with the blue cloud.
\begin{figure*}[t!]
    \centering
    \includegraphics[width=1\textwidth]{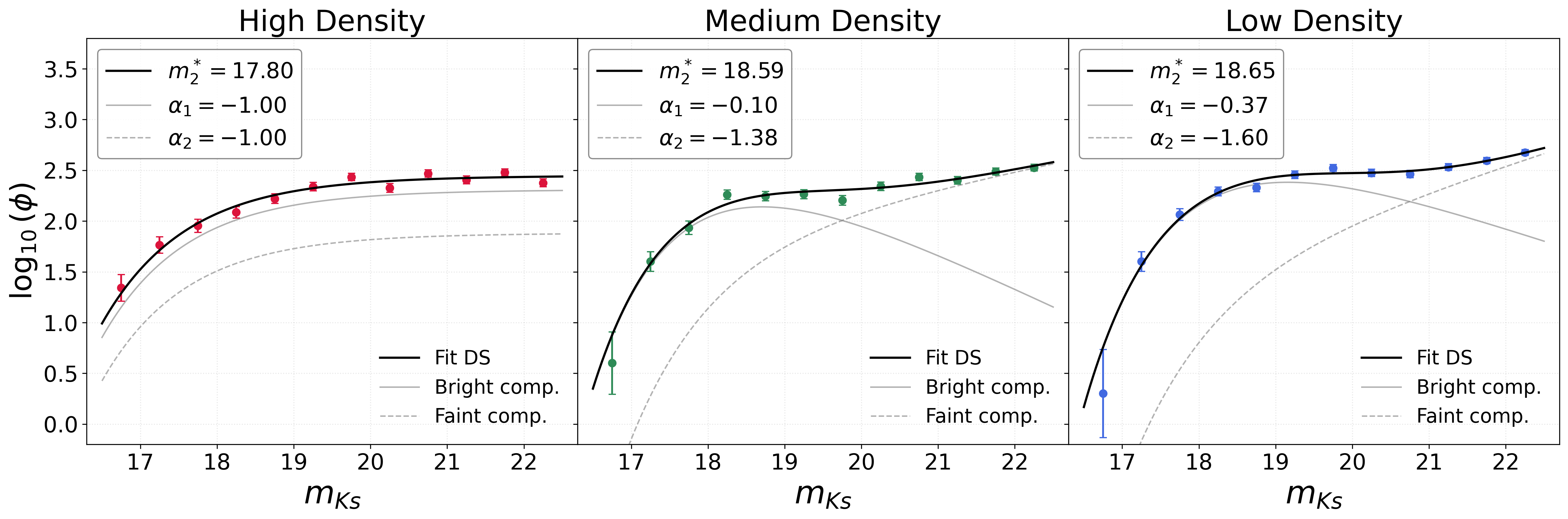}  
    \caption{\small $Ks$-band luminosity functions for galaxies in high (left, red), medium (center, green), and low (right, blue) density environments. Observed number densities ($\log_{10} \phi$) are plotted with 1$\sigma$ Poissonian error bars. Solid black lines represent the best-fit Double Schechter functions, while dotted and dashed lines show the bright and faint components, respectively. The lower axis indicates the corresponding apparent $Ks$ magnitudes assuming a distance modulus of the cluster at $z=$0.346.}
    \label{fig:lf}
\end{figure*}

\subsection{The Ks-band Luminosity Function}
\label{sec:5.4}
The Luminosity Function \citep[LF, ][]{Schecter76} represents a fundamental tool for quantifying the distribution of galaxy populations and their evolution within different environments. In this work, we express the LF in terms of apparent magnitude $m_{Ks}$, considering the combined sample of spectroscopic and photometric cluster members identified in Sec.~\ref{sec:4.2}. By dividing all cluster members into three local density ranges, we quantify the distribution of galaxies as a function of their luminosity, providing a statistical representation of the galaxy population in different environments. We describe the observed LFs using a Double Schechter function (DS). Although the fit of a single Schechter function cannot be statistically rejected in all density bins, the observed distributions, especially in the lower density regions, are better described by a sum of two Schechter functions, representing the bright and faint galaxy populations, respectively. Following the approach of \citet{Baldry08, Peng10}, we formalize this description using a double Schechter function (see Eq.~\ref{eq:double_schechter_ks}):

\begin{equation}
\begin{split}
\Phi(m_{Ks}) \, \mathrm{d}m_{Ks} = \Bigl[ & \phi_1^* 10^{0.4(m^* - m_{Ks})(\alpha_1 + 1)} \\ 
& + \phi_2^* 10^{0.4(m^* - m_{Ks})(\alpha_2 + 1)} \Bigr] e^{-10^{0.4(m^* - m_{Ks})}} \mathrm{d}m_{Ks},
\end{split}
\label{eq:double_schechter_ks}
\end{equation}
\\where $m_{Ks}$  is the apparent magnitude in the $Ks$ band and $m^*$=17.8 mag is value derived from the initial single Schechter fit as a fixed reference point to ensure fit stability and reduce parameter degeneracy across different density bins. In this parametrization, $\alpha_1$ and $\alpha_2$ represent the slopes of the two Schechter components: $\alpha_1$ describes the spectral index of the bright component (massive galaxies), dominating the "knee" of the distribution, while $\alpha_2$ represents the faint component (dwarf galaxies), responsible for the steepening slope observed at the faintest magnitudes. The $\phi_1^*$ and $\phi_2^*$ parameters are the respective normalization constants. The best-fit parameters (see Table~\ref{tab:LF_params}) were derived using a Maximum Likelihood estimation based on Poisson statistics, following the approach often adopted in cluster studies \citep[][]{Malumuth86}. 
The evolution of the Schechter parameters reveals a clear environmental dependence:
\begin{itemize}
    \item \textbf{High Density}: in the densest regions, the distribution is effectively described by a nearly single Schechter, as the two components converge toward an identical slope ($\alpha_1 = \alpha_2 = -$1.01). This confirms that the cluster core is dominated by a relatively homogeneous population of massive, passive galaxies. The characteristic magnitude $m_2^* = $17.80 represents the bright luminosity scale of these dominant systems, while the faint dwarf population appears significantly suppressed;
    \item \textbf{Medium Density}: in intermediate regions, the characteristic magnitude shifts to a fainter value of $m_2^* = $18.59. We observe a flattening of the bright slope to $\alpha_1 = -$0.10, while the faint slope reaches $\alpha_2 = -$1.38;
    \item \textbf{Low Density}: in the lowest density regions, the LF exhibits a remarkably steep faint slope ($\alpha_2 = -$1.60), indicating a dominant population of dwarf galaxies. The characteristic magnitude shifts further toward fainter values ($m_2^* = $18.66) and a visible "dip" in the distribution confirms the presence of two distinct populations, massive galaxies and a well preserved population of fainter, star-forming blue galaxies which are effectively captured by the DS model ($\alpha_1 = -$0.37).
\end{itemize}

\begin{table}[h]
\centering
\begin{tabular}{lccc}
\hline
Environment & $m_2^*$ & $\alpha_1$ & $\alpha_2$ \\
density & [mag] & (bright slope) & (faint slope) \\
\hline
High   & 17.80 $\pm$ 0.14 & $-$1.00 $\pm$ 0.10 & $-$1.00 $\pm$ 0.24 \\
Medium & 18.59 $\pm$ 0.32 & $-$0.10 $\pm$ 0.59 & $-$1.38 $\pm$ 0.20 \\
Low    & 18.65 $\pm$ 0.19 & $-$0.37 $\pm$ 0.33 & $-$1.60 $\pm$ 0.36 \\
\hline
\end{tabular}
\caption{Best-fit parameters of the DS $Ks$-band LF for the three density regimes. The magnitude $m_2^*$ of the second Schechter, the slopes $\alpha_1$ (bright component) and $\alpha_2$ (faint component) are reported with their $1\sigma$ uncertainties derived from the numerical Hessian matrix.}
\label{tab:LF_params}
\end{table}
This trend is consistent with a scenario where environmental mechanisms, such as mergers, tidal stripping, and quenching \citep{Annunziatella16, Chan19, Stephenson25}, actively shape the galaxy population. In the high-density core, these processes lead to the disruption or assimilation of dwarf galaxies, effectively smoothing the LF into a single-component profile. Conversely, as we move toward the outskirts (low density), the systematic shift of $m^*$ toward fainter magnitudes and the steepening of $\alpha_2$ confirm that low-mass systems are not only more frequent but are better preserved in less extreme environments.
\section{Summary and discussion}
\label{sec:6}
The results presented in this paper emphasize the importance of extending galaxy evolution studies up to 5$r_{200}$ to capture the full diversity of cluster environments. By leveraging the unique combination of high-resolution VST optical data and NIR VISTA imaging, we performed a multiwavelength photometric analysis of the galaxy cluster AS1063 at $z = $0.346. This synergy enabled the construction of a robust catalog (described in Section~\ref{sec:3}) containing 64394 objects, reaching a completeness limit of $r < $24.65 mag. The primary advantage of investigating the cluster outskirts at such large radial distances lies in the ability to characterize infalling galaxy populations before they are fully processed by the extreme conditions of the cluster core. This wide-field approach allows us to map galaxy properties, from the high-density central regions to the low-density outskirts, providing a comprehensive view of the environmental mechanisms driving galaxy transformation at an epoch of rapid evolution.
\\ In~\ref{sec:4.2}, we determine photometric redshifts for all galaxies in our catalog, using a machine learning approach based on the Random Forest method. Spectroscopic redshifts available in the central region of the cluster were employed to calibrate the SED fitting procedure and to assess the accuracy of the resulting photometric estimates. The derived photometric redshifts are robust, characterized by a bias of 0.006, a smoothing scale $\sigma = $0.10, a scatter $\sigma_{\text{NMAD}} = $0.02, and an outlier fraction $\eta =$ 6.2$\%$. Figure~\ref{fig:range} show the photometric redshift range 0.304 $ \leq z_{\text{spec}} \leq $ 0.408 in which we identified 3693 photometric galaxy members. Accounting for the spectroscopic members, the final sample consists of 4927 cluster galaxies. We estimated a completeness of 90.7$\%$ and a purity of 69.7$\%$ for this selection. Utilizing this comprehensive membership, we performed a fit of the red sequence (see Fig.~\ref{fig:red}). This reliable membership assignment provides the necessary statistical foundation to investigate galaxy evolution across the diverse density regimes of AS1063.
\\ A key result of this study is the reconstruction of the density field (see Fig.~\ref{fig:density}), which reveals a complex web of structures extending up to 5$r_{200}$. This wide-field analysis allowed us to identify features that are typically inaccessible in standard cluster studies, most notably a dense clump located 11.26 Mpc from the center in the NW direction. This clump is connected to the cluster core by a prominent filamentary structure. The detection of these filaments is crucial, as they provide the physical pathways for infalling galaxy populations and offer a unique laboratory to study pre-processing mechanisms far from the virial radius.
\\The analysis of the Color-Magnitude Diagrams (see Fig.~\ref{fig:redsequence}) provides a clear visual result of the environmental dependence of galaxy colors. Our results show that while the high-density core ($\delta_{smooth}>$2.3) is dominated by a well populated red sequence of 1333 galaxies, the population of blue galaxies is almost exclusively found in the lower-density outskirts ($\delta_{smooth}<$1.2). This trend confirms that blue, star-forming galaxies are well-preserved in the cluster periphery and along the filaments, but are efficiently transformed into red, passive systems as they fall toward the central regions. This spatial segregation highlights the crucial role of the environment in quenching star formation during cluster assembly.
\\ The last result is visible in Fig.~\ref{fig:lf}, where we show the $Ks$-band LF that provides a quantitative measure of how galaxy mass distributions shift with density. We fit a Double Schechter function and fixing the value of the magnitude of the first component ($m_1^* = $17.8 mag). In high-density regions, the two components of the fit converge toward an identical slope ($\alpha_1 = \alpha_2 = -$1.01), indicating that the cluster core is a mass-segregated environment where massive galaxies dominate, while the low-mass population has been significantly suppressed. In contrast, the low-density LF exhibits a prominent "dip" and a strikingly steep faint-end slope ($\alpha_2 = -$1.60), with the characteristic magnitude shifting toward fainter values ($m^* = $18.66). This mass-dependent trend confirms that while massive galaxies dominate the cluster center, the lower-density outskirts better preserve the infalling dwarf population, which is otherwise disrupted or transformed by mechanisms such as tidal stripping and mergers upon entering the high-density cluster environment.
\\ The multiwavelength photometric catalog derived in this study is released with this paper. For access to the individual catalogs for each band, researchers are invited to contact the author. Looking forward, we plan to extend this analysis by targeting the newly identified overdense clumps and filamentary structures with follow-up spectroscopic observations to confirm their dynamical state and membership. These future works will focus on a detailed comparison of galaxy properties across the different density regimes identified here, with a specific emphasis on the role of galaxy stellar mass. By investigating the mass-dependent evolution and the impact of pre-processing in the cluster outskirts, we aim to further clarify the fundamental mechanisms that drive the transformation of galaxies as they move from the field into the high-density cluster environment.

\begin{acknowledgements}
Aniello Grado acknowledges the support of the Italian Ministry of University and Research (MUR) through the program “Dipartimenti di Eccellenza 2023–2027” (Grant SUPER-C). PB acknowledges financial support through grant PRIN-MIUR 2020SKSTHZ and support from the Italian Space Agency (ASI) through contract "Euclid - Phase E", INAF Grants "The Big-Data era of cluster lensing" and "Probing Dark Matter and Galaxy Formation in Galaxy Clusters through Strong Gravitational Lensing". 
\end{acknowledgements}

\bibliographystyle{aa}
\bibliography{bibliografia}

\clearpage
\appendix
\section{Masking procedure}
\label{app:A}
\begin{figure}[h]
    \centering
    \includegraphics[width=1.0\linewidth]{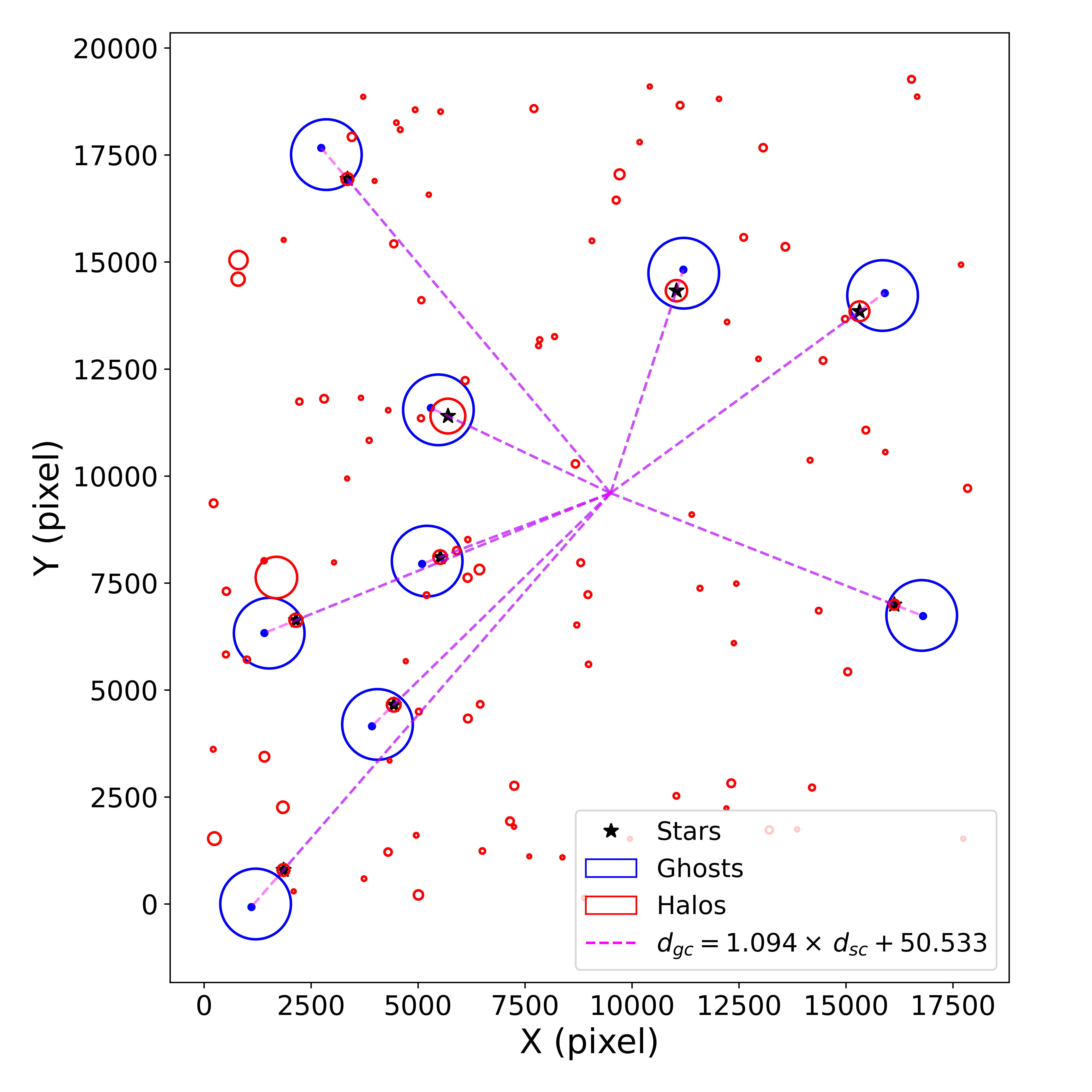}
    \caption{\small Masking for the $r$-band. We show the halos as red circles and the ghosts as blue ones. The stars that produced these artifacts are in black. The relation between the distance center-stars and center ghosts are also reported: $d_{gc} = 1.094 \times d_{sc} + 50.533$.}
    \label{fig:masking}
\end{figure}

In this appendix, we explain in more detail the masking procedure used for halos and ghosts for VST and VISTA images. We use the same semi-automatic procedure for each band as already done in \citet{Estrada23}.
The halos are centered on the stars that generate them, but vary according to the magnitude of the star itself. In Tables~\ref{tab:mask_VST} and~\ref{tab:mask_VISTA}, we report the magnitude ranges, with steps of one magnitude, and the relative halo radii for each band. We note that there is a linear relationship between the magnitude and the halo radii: as the magnitude increases, the radius increases by 0.5 \texttt{FLUX\_RADIUS\_90} (R$_{90}$) defined as the radius containing 90$\%$ of the total flux of the object. With a visual approach, we verify for each band at which magnitude value the stars no longer produce halos. For the $u$ band, we used only stars listed in the Gaia catalog, since the objects producing halos and ghosts are not detected by SExtractor. Consequently, the halos are generated using a handmade procedure using a physical size of 450. Indeed, for the ghosts we use the same approach in all bands.
\begin{table}[h]
\centering
\begin{tabular}{lccccc} 
\toprule 
 Gaia range & g radius & r radius & i radius \\   
$[\mathrm{mag}]$ & [$\times R_{90}$] & [$\times R_{90}$] & [$\times R_{90}$] \\
        \midrule 
         $G\leq 8.5$ & 1.5 & 1.0 & 1.0 \\ 
         $8.5<G\leq 9.5$ & 2.0 & 1.5 & 1.5 \\ 
         $9.5<G\leq 10.5$ & 2.5 & 2.0 & 2.0 \\ 
         $10.5<G\leq 11.5$ & 3.0 & 2.5 & 2.5 \\ 
         $11.5<G\leq 12.5$ & 3.5 & 3.0 & 3.0 \\ 
         $12.5<G\leq13.5$ & - & 3.5 & - \\ 
        \bottomrule 
    \end{tabular}
    \caption{\small Gaia magnitude range to detect halos generated by the brightest stars in the optical bands. The saturation limit is 99.0 mag for $u$ band, 13.0 for the $g$ band, 13.9 mag for $r$ band and 12.9 mag for $i$ band.}
    \label{tab:mask_VST}
\end{table}
\begin{table}[h]
\centering
\begin{tabular}{lccc} 
\toprule 
Gaia range & Y radius & J radius & Ks radius  \\   
$[\mathrm{mag}]$ & [$\times R_{90}$] & [$\times R_{90}$] & [$\times R_{90}$] \\
    \midrule 
         $G\leq 7.5$ & 4.0 & - & 3.5  \\
         $7.5<G\leq 8.5$ & 4.5 & 3.5 & 4.0 \\ 
         $8.5<G\leq 9.5$ & 5.0 & 4.0 & 4.5 \\ 
         $9.5<G\leq 10.5$ & 5.5 & 4.5 & 4.5 \\ 
         $10.5<G\leq 11.5$ & - & 5.0 & 4.5 \\ 
         $11.5<G\leq 12.5$ & - & 5.5 & - \\ 
        \bottomrule 
    \end{tabular}
    \caption{\small Gaia magnitude range to detect halos generated by the brightest stars in the NIR bands. The saturation limit is 11.6 mag for $Y$-band, 13.2 for the $J$-band and 13.2 mag for $Ks$-band.}
    \label{tab:mask_VISTA}
\end{table}
\\ Unlike halos, ghosts are not centered on the star that produces them but are more displaced from the center of the star the farther they are from the center of the field, in the radial direction. Once the center of each image is identified, we use an automatic procedure that generates a mask for the VST and VISTA field of view based on radial criteria. First, we visually select some ghosts generated by the brightest stars and note that their size is the same for all optical bands and varies for NIR images, but remains the same within the same band as reported in Table~\ref{tab:mask2}. We find a linear relation for the distance d$_{SG}$ star-hand-made ghost as a function of the distance d$_{SC}$ star-center of the field: d$_{SG}$ = $a \times d_{SC} + b$ measured in pixels. We verify that each star is associated with only one ghost and that the radial criterion is respected. Using this relation, we automatically generate masks for all stars that produce them. In Fig.~\ref{fig:masking}, we show the masking for $r$ band, in which there are halos (red) and ghosts (blue) produced by saturated stars (black). 
\begin{table}[h]
\centering
\begin{tabular}{lccccc} 
\toprule 
 Band & Gaia limit  & a & b & ghosts radius  \\
     &  [mag] & [pixel] & [pixel] & [pixel] \\
        \midrule 
     $u$  &   $G\leq 10.3$ & 1.094 & 48.404  & 825  \\
     $g$  &   $G\leq 11.0$ & 1.062 & 90.035  & 825  \\
     $r$  &   $G\leq 11.5$ & 1.094 & 50.533  & 825  \\
     $i$  &   $G\leq 11.4$ & 1.087 & 16.579 & 825  \\
     $Y$  &   $G\leq 11.5$ & 1.011 & 22.420  & 400  \\
     $J$  &   $G\leq 12.9$ & 1.022 & 3.1180  & 550  \\
     $Ks$ &   $G\leq 9.1$  & 1.010 & 19.088  & 400  \\
        \bottomrule 
    \end{tabular}
    \caption{\small The Gaia magnitude limit of stars that produced ghosts for each band. The parameter a and b are related, respectively, to the distance from the generating ghost $d_{gc}$  and from the star $d_{sc}$ to center of images. The radius, in pixel, is the size of the ghosts for each band.}
    \label{tab:mask2}
\end{table}
\begin{table}[h]
\centering
\begin{tabular}{ccccccc} 
\toprule 
 Bands & N sources & $\%$ sources & N sources & $\%$ sources \\   
   &  no-masked & no-masked & masked & masked \\
        \midrule 
         $u$ & 6039 & 2.7$\%$ & 221925 & 97.3$\%$\\
         $g$ & 22283 & 9.7$\%$ & 208198 & 90.3$\%$\\ 
         $r$ & 29269 & 12.8$\%$ & 198695 &  87.2$\%$\\ 
         $i$ & 14353 & 13.3$\%$ & 93331 & 86.7$\%$\\ 
         $Y$ & 6546 & 2.5$\%$ & 256742 & 97.5$\%$\\ 
         $J$ & 34850 & 10.5$\%$ & 297257 & 89.5$\%$\\ 
         $Ks$ & 1940 & 0.8$\%$ & 241897 & 99.2$\%$\\ 
        \bottomrule 
    \end{tabular}
    \caption{\small Masked source numbers and their percentages for all bands.}
    \label{tab:mask_source}
\end{table}
\\ Spikes are manually identified in all bands, since no correlation is observed between their length and the Gaia G magnitude of the stars generating them.
\\ Finally, after the masking procedure we report in Table~\ref{tab:mask_source} the number of sources inside (\texttt{FLAG\_MASK}$=$0) and outside the masks (\texttt{FLAG\_MASK}$=$1, \texttt{FLAG\_MASK}$=$2, \texttt{FLAG\_MASK}$=$3 if are in halos, ghosts and spikes, respectively).

\section{Appendix B: JWST data reduction}
\label{app:B}
The reduction of JWST/NIRCam observations was carried out starting from the raw data to produce a final science-ready mosaic. The available NIRCam observations of AS1063 (PI: Atek H., program GO 3293) were retrieved from the MAST archive\footnote{\url{https://mast.stsci.edu/search/ui/##/}}. The program includes observations for the following bands: F090W, F115W, F150W, F200W, F277W, F410M, F356W, F410M, F444W, F480M. In this Appendix, we describe how the individual exposures were processed, aligned, and combined into the final mosaic. We will use F200W to calibrate the NIR $Ks$ band. For the reduction, we used a combination of the JWST pipeline \citep[][]{JWST16}, the Bagley reduction \citep[][]{Bagley23}, and custom scripts. This dataset is quite large: there are 384 exposures for each NIRCam module (768 in total). Following JWST official pipeline, data reduction is performed in three stages. Throughout the process we used the pmap: jwst\_1322.pmap.
\\ The Stage 1 of JWST data reduction is to transform the raw files downloaded from the MAST archive, those ending in \textit{uncal.fits}, into the first-level images with units of DN/s (Data Number/sec). This step is crucial, as it removes most instrumental effects through detector-level corrections, many of which are in common to all instruments and observing modes. Stage 1 converts the raw detector signals into a physically meaningful form by i) initializing pixel quality masks; ii) correcting for saturation effects and the nonlinear response of the detector; iii) subtracting the superbias and dark current; iv) removing the contributions of the reference pixels; v) identifying and marking cosmic rays; vi) performing 1/f noise correction; vii) combining successive readings through ramp fitting to produce a final image calibrated in counts per second. The output of Stage 1 consists of 2D and 3D detector images, stored in the extension \textit{rate.fits}, ready to be processed in Stage 2, where astrometric and flux calibration is performed.
\\ Stage 2 produces scientifically calibrated images as output. During this Stage, each individual image is processed by applying astrometric calibration and additional optical or instrumental corrections. The images are mapped onto celestial coordinates (RA, Dec) using the updated WCS solution, aligned to ensure astrometric consistency across all exposures, corrected for small positional errors and residual artifacts, and calibrated in flux units (MJy/sr). The output of Stage 2 consists of individually calibrated and astrometrically aligned images \textit{cal.fits}, in units of MJy/sr. At this stage, the exposures are fully corrected for instrumental effects, their WCS solution is refined using the Gaia catalog as astrometric reference, and they are flux-calibrated and ready to be used individually for scientific analysis.
\\ These two reduction stages are performed independently for each exposure.
\\ Stage 3 combines all calibrated exposures from the same filter into a final science-ready mosaic. In this stage, residual outliers, such as defective pixels and cosmic rays not fully removed in Stage 1, are identified and flagged. The calibrated images are then further refined through two-dimensional background modeling and subtraction, ensuring a consistent sky level across the field, while preserving extended and low-surface-brightness features. The WCS solution is verified and updated when necessary, and the associated variance map is adjusted to provide realistic noise estimates and proper weighting during images combination. The final mosaic is produced using calwebb\_image3 from pipeline, which resamples and co-adds the aligned exposures onto a common grid. The resulting product is a flux-calibrated mosaic with updated variance information, delivered as an \textit{i2d.fits} file.

\section{Appendix C: photometric redshift}
\label{app:C}
Photometric redshifts were estimated in the first instance using the EAZY software \citep{Brammer08}, a standard template-fitting approach. However, this method yielded suboptimal results for our dataset, specifically achieving a completeness of 65.56$\%$ and a purity of 57.1$\%$. Such values were considered insufficient for a robust analysis, as they would introduce significant background contamination and the loss of numerous faint members. Consequently, we implemented a Machine Learning-based approach, which significantly improved the membership selection, providing a more reliable classification for the cluster population
\\ Machine learning (ML) paradigms have been widely proven to be a powerful alternative to spectral energy distribution (SED) fitting methods for estimating photometric redshifts {\citep[][]{Cavuoti12, Cavuoti15, Tucci25}. We tested three supervised ML regressors and compared their performances in predicting the photometric redshifts: a K-Nearest Neighbors (KNN, \citet{Fix89} and \citet{Cover67}, a Random Forest \citet{Ho98}, and a Multi-Layer Perceptron \citep[][]{Bishop07} implemented with a Quasi-Newton Algorithm (MLPQNA, as described in \citet{Brescia19} and \citet{Cavuoti17}.
\\Supervised ML paradigms need to be trained on a dataset, the so-called knowledge base (KB), that contains a set of input features and the target quantity one is trying to predict. During the training phase, the ML models learn the complex, hidden relations that link the input features to the target feature, which is the spectroscopic redshift in our case. We built our KB using a subset of 2514 objects in the photometric catalog that have a spectroscopic redshift $z_{\text{spec}}$, that are not stars ($z_{\text{spec}} \geq $0.01), have no invalid Kron or aperture magnitude value in any of the available photometric bands, and errors on magnitudes lower than 0.5. We then randomly split the KB into two datasets used for training and validation, containing 1058 ($\sim $60$\%$) and 1006 ($\sim$ 40$\%$) objects, respectively. Following standard practice in ML applications, the training dataset was standardized by subtracting the median of each input feature and dividing by the corresponding inter-quartile range (IQR)\footnote{The IQR is defined as the difference between the 75th (Q3) and 25th (Q1) percentiles of the data distribution.}. The median and IQR values computed on the training dataset are then used to standardize the validation datasets. We trained the three ML regressors on the training dataset to predict $z_{\text{spec}}$, using the \texttt{MAG\_AUTO} magnitudes in the seven available photometric bands as input features. We then assessed the performance of the trained regressors by predicting the $z_{\text{phot}}$ for the objects in the validation dataset (which had never been presented to the models during the training phase) and comparing them with the respective $z_{\text{spec}}$. The three regressors showed a performance comparable to eazy in terms of $|\text{bias}|$, $\delta_{smooth}$, $\sigma_{\text{NMAD}}$, and $\eta$, with both the KNN and RF having a lower value of $\sigma \simeq $0.15 and only the MLPQNA had a slightly higher value of $\sigma_{\text{NMAD}} = $0.24. We also computed P and C of the photometric cluster members identified using the photometric redshift range 0.304$ \leq z_{\text{phot}} \leq $0.408. Both the MLPQNA and KNN reached a higher completeness than Eazy, but with a lower purity, while the RF performed better on both metrics, reaching $\text{P} = $67.5$\%$ and $\text{C} = $81.6$\%$. We report the values of the statistical indicators in Table\ref{tab:regressor_validation_statistics}, to make the comparison among the models and the different configurations easier.
\\ To improve the performance of the model, we repeated the training and validation process, including as input features several aperture magnitudes in addition to the Kron ones, together with their corresponding natural colors (namely, $u-g$, $g-r$, $r-i$, etc.). We found that the performance of MLPQNA improved when magnitudes and colors from a single additional aperture were included. Conversely, the performance of KNN and RF showed no significant improvement when using magnitudes and colors extracted from apertures larger than $3''$ (MAG\_AP3). The validation indicated that the best results for MLPQNA were obtained with the combination of only Kron and $3''$ aperture magnitudes and colors, for a total of 27 input features, yielding $|\text{bias}| = $0.03, $\sigma = $0.18, $\sigma_{\text{NMAD}} = $0.04, and $\eta = $11.1$\%$, with a purity $\text{P} = $65.2$\%$ and completeness $\text{C} = $78.3$\%$. For the KNN and RF regressors, the best performance was achieved by using all the Kron, $1.5''$ (MAG\_AP1), $2''$ (MAG\_AP2), and $3''$ aperture magnitudes and colors, for a total of 52 input features. In both cases, this input feature set outperformed the results obtained with eazy as well as those from the previous setups, especially in terms of catastrophic outliers fraction, $\delta_{smooth}$, and $\sigma_{\text{NMAD}}$. In particular, the KNN achieved $|\text{bias}| = $0.003, $\sigma = $0.12, $\sigma_{\text{NMAD}} = $0.04, and $\eta = $5.8\%, reaching a completeness $\text{C} = $81.5$\%$ but with a lower purity $\text{P} = $57.8$\%$, indicating that there is a high contamination from non member galaxies falling in the $z_{\text{phot}}$ membership range, despite the overall high accuracy in predicting the redshift. The RF, instead, achieved $|\text{bias}| = $0.007, $\sigma = $0.10, $\sigma_{\text{NMAD}} = $0.02, and $\eta = $6.2$\%$, reaching a high level of completeness $\text{C} = $91.0$\%$ and a purity $\text{P} = $69.9\% (see Table \ref{tab:regressor_validation_statistics}).
\begin{table*}[!ht]
    \centering
    \begin{tabular}{l c ccc ccc c}
        \toprule
        Input features & Kron mag.s & \multicolumn{3}{c}{Kron mag.s} & \multicolumn{3}{c}{Kron and aperture mag.s and col.s} & Best feature subset \\
                   \cmidrule(lr){2-2} \cmidrule(lr){3-5}            \cmidrule(lr){6-8}           \cmidrule(lr){9-9}
        and model   &  EAZY   &  MLPQNA &   KNN   &    RF   &  MLPQNA &   KNN   &   RF   &     RF  \\ 
        \midrule
    $|\text{bias}|$ &  0.001  &    0.019  &  0.001  &  0.013  &  0.03   &  0.003  & 0.007  &   0.006  \\
           $\sigma$ &  0.21   &    0.24   &  0.15   &  0.14   &  0.18   &  0.12   & 0.10   &   0.10   \\
    $\sigma_{NMAD}$ &  0.05   &    0.05   &  0.05   &  0.04   &  0.04   &  0.04   & 0.02   &   0.02   \\
             $\eta$ &  8.7\%  &    9.2\%  &  8.7\%  &  8.7\%  & 11.1\%  &  5.8\%  & 6.2\%  &   6.2\%  \\
                  P & 57.1\%  &   60.3\%  & 58.9\%  & 67.5\%  & 65.2\%  & 57.8\%  & 69.9\% &  69.7\%  \\
                  C & 65.6\%  &   66.1\%  & 74.8\%  & 81.6\%  & 78.3\%  & 81.5\%  & 91.0\% &  90.7\%  \\
        \bottomrule
    \end{tabular}
    \caption{Statistical indicators for the different photometric redshift estimators and configurations tested. EAZY statistics have been computed on the entire spectroscopic dataset, while the one for the three ML models have been computed on a validation dataset that contains objects not used for the training phase.}
    \label{tab:regressor_validation_statistics}
\end{table*}
\begin{figure*}
    \centering
    \includegraphics[width=1.0\linewidth]{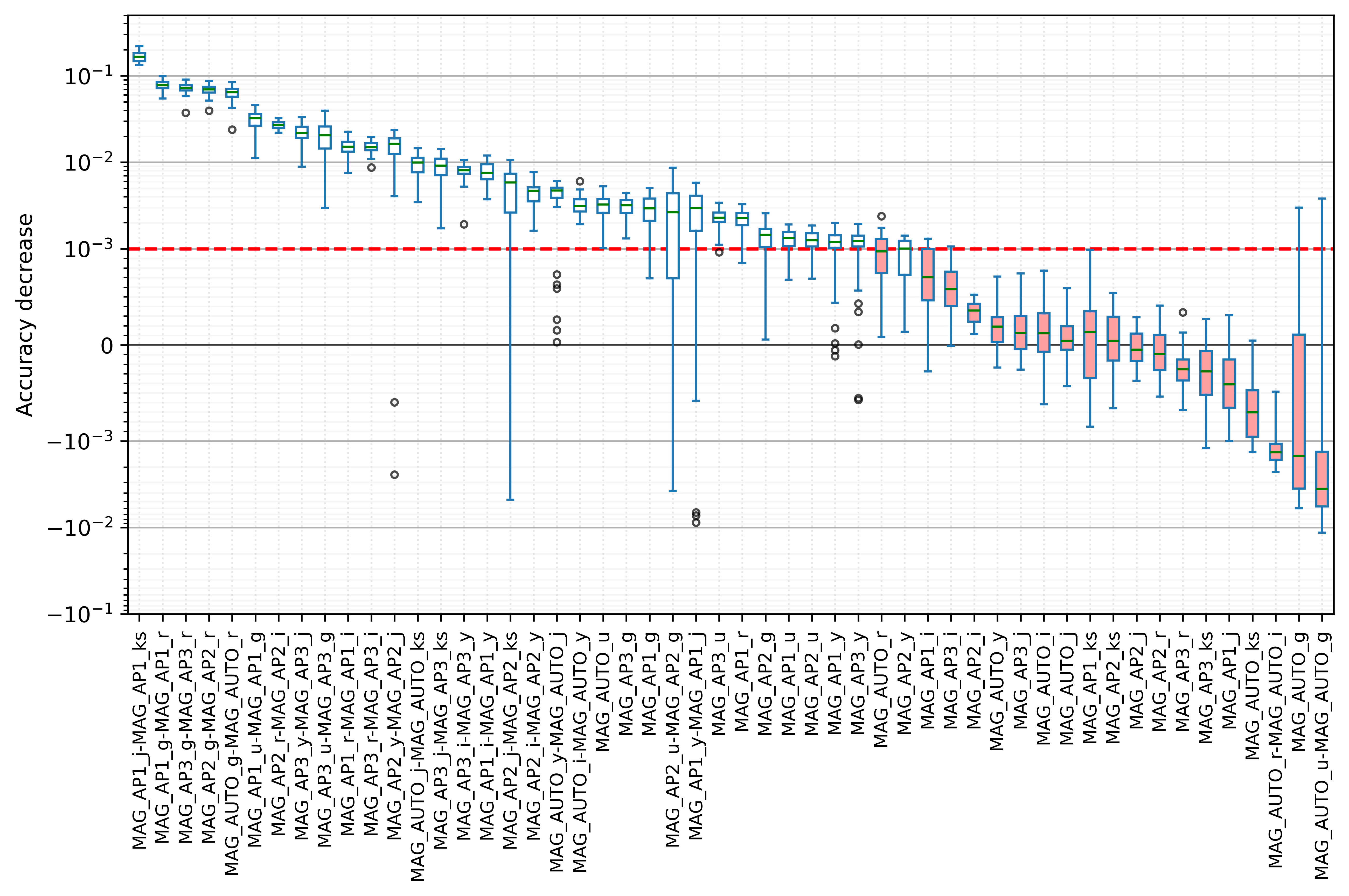}
    \caption{Permutation feature importance for the RF regressor. The permutation importance was computed 50 times for each of the 52 input features: the green line indicates the median value of the accuracy decrease; the vertical boxes indicate the first and the third quartile; the whiskers mark the percentiles that correspond approximately to $\pm 3 \sigma$; the black circles indicate outliers outside the $\pm 3 \sigma$ range. The 19 features that were excluded in the feature selection process are marked in red, and the red dashed line indicates the $0.1\%$ accuracy decrease threshold. The vertical axis follows a symmetric logarithmic scale, with the linear zone between $-10^{-3}$ and $10^{-3}$.}
    \label{fig:rf_fi}
\end{figure*}
\\Since the RF regressor showed a higher accuracy in predicting the $z_{\text{phot}}$, together with a high value of both P and C of the selected cluster members, we focused the subsequent optimization exclusively on this model. In particular, we computed the permutation importance \citep[][]{Breiman01} of each input feature using the implementation available in the scikit-learn Python library, and retained only the most relevant ones. The permutation importance is a model-agnostic technique commonly used to estimate how much each input feature contributes to a model’s predictive performance \citep[][]{Fisher18}. In our case, the importance of each feature is measured by first computing the root mean squared error ($\text{RMSE}_{\text{orig}}$) between the $z_{\text{spec}}$ and the $z_{\text{phot}}$ predicted with the RF on the validation dataset. Then, the values of a single feature were randomly shuffled across the validation dataset, preserving its marginal distribution but breaking the relation with the $z_{\text{spec}}$. The RF is then used to compute the root mean squared error ($\text{RMSE}_{\text{perm}}$) using this modified validation dataset and the $\text{RMSE}_{\text{perm}}$, and the ratio $\text{RMSE}_{\text{perm}}/\text{RMSE}_{\text{orig}}$ provides an estimation of the accuracy decrease of the regressor due to the loss of information in the shuffled feature (the greater is the accuracy loss, the more important is the shuffled feature). This process is repeated 50 times for each feature, and we discarded 19 input features that show a median accuracy decrease lower than the 0.1$\%$ (see Fig.\ref{fig:rf_fi}). Of the 33 selected "best" input features, there are: 22 Kron and aperture natural colors (only the Kron $u-g$ and $r-i$ were excluded); 10 aperture magnitudes in the bands $u$, $g$, $r$, and $Y$; and the Kron magnitude in band $u$.
\\ One drawback of the permutation importance is that it is particularly affected by collinearity among input features. In fact, shuffling one correlated feature may not strongly degrade model accuracy if it can still rely on the information that is provided by the correlated features, and could lead to the pruning of all the collinear features, since they will all appear not important. Therefore, we verified that for the magnitudes in the $i$, $J$, and $Ks$ bands, which were completely discarded, there was at least one collinear feature in the best input features subset. We computed the correlation matrix of all the input features and, from this, the Ward's linkage distance \citep[][]{Ward1963}, which can be used to cluster together similar features (see Fig. \ref{fig:rf_fi}). Both the correlation matrix and the clustering dendrogram show that the magnitudes in the $J$ and $Ks$ bands are highly correlated with those in the $Y$ band. Similarly, the magnitudes in the $i$ band are highly correlated to those in the $r$ band. Since in the "best" input feature subset there is at least one magnitude from the $r$ and $Y$ bands, no information is lost during the feature selection. This is also confirmed by the fact that there is no actual loss in performance when training an RF using only the "best" selected features (see Table \ref{tab:regressor_validation_statistics} and Fig.\ref{fig:ref_fi_zspec_vs_zphot}). We then used the RF trained using only the "best" input features to predict the $z_{\text{phot}}$ for the objects in the photometric catalog that have no spectroscopic redshift.
\begin{figure}
    \centering
    \includegraphics[width=\linewidth]{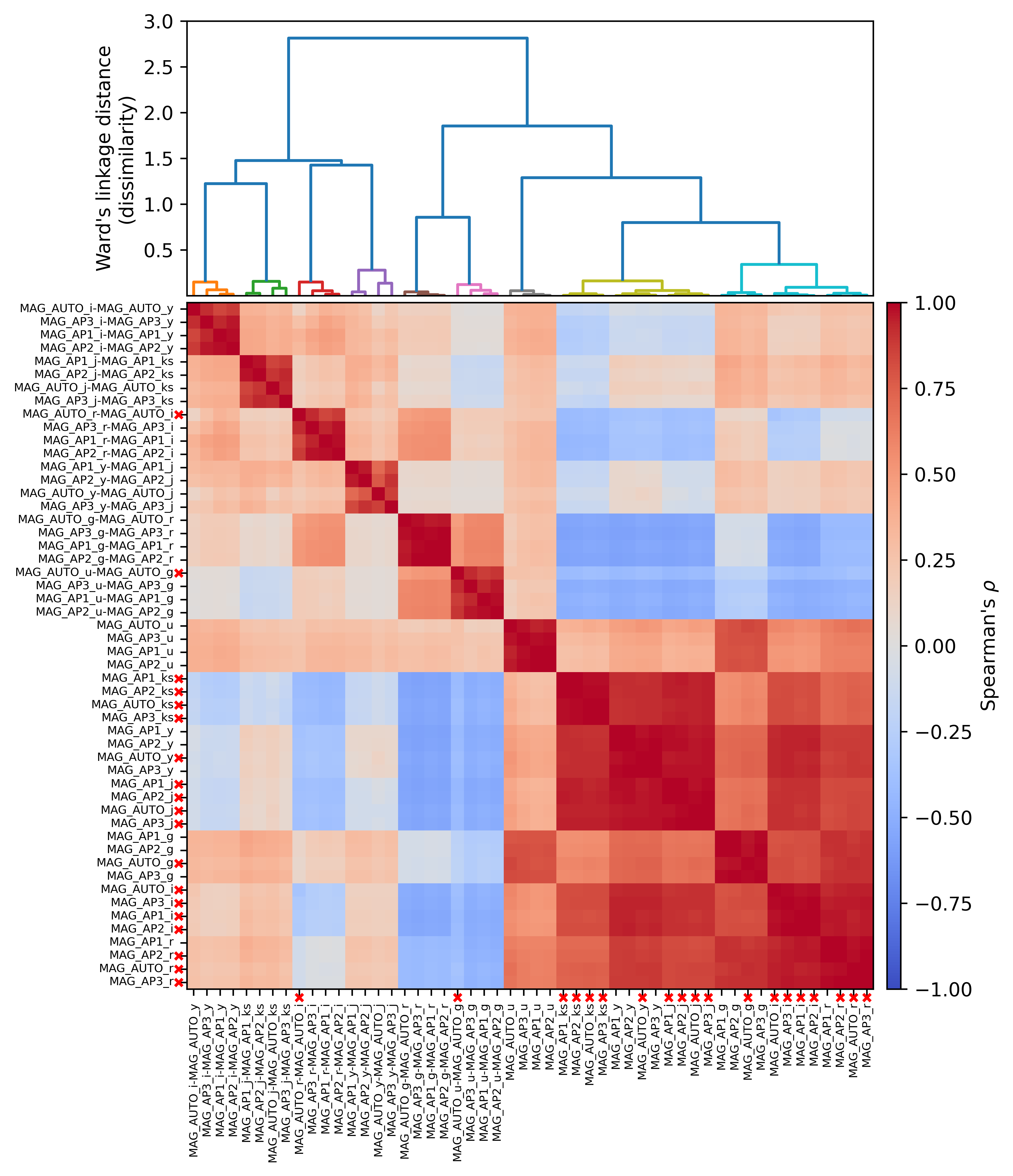}
    \caption{Input features correlation matrix (lower panel) and Ward's linkage distance dendrogram (upper panel). The correlation matrix shows the Spearman's rank correlation coefficient $\rho$ for each possible pair of input features, where red indicates $\rho = $1 (perfect collinearity), blue represents $\rho = -$1 (perfect anti-linearity), and gray represents $\rho = $0 (no correlation). The Ward's linkage distance dendrogram shows how similar input features are clustered together according to the Ward's method \citep[][]{Ward1963}. Features that are closer than 0.5 in Ward's distance are clustered together and have the same color in the dendrogram. The red crosses indicate the features that were excluded in the feature selection.}
    \label{fig:rf_cm_and_dendrogram}
\end{figure}
\begin{figure}
    \centering
    \includegraphics[width=0.9\linewidth]{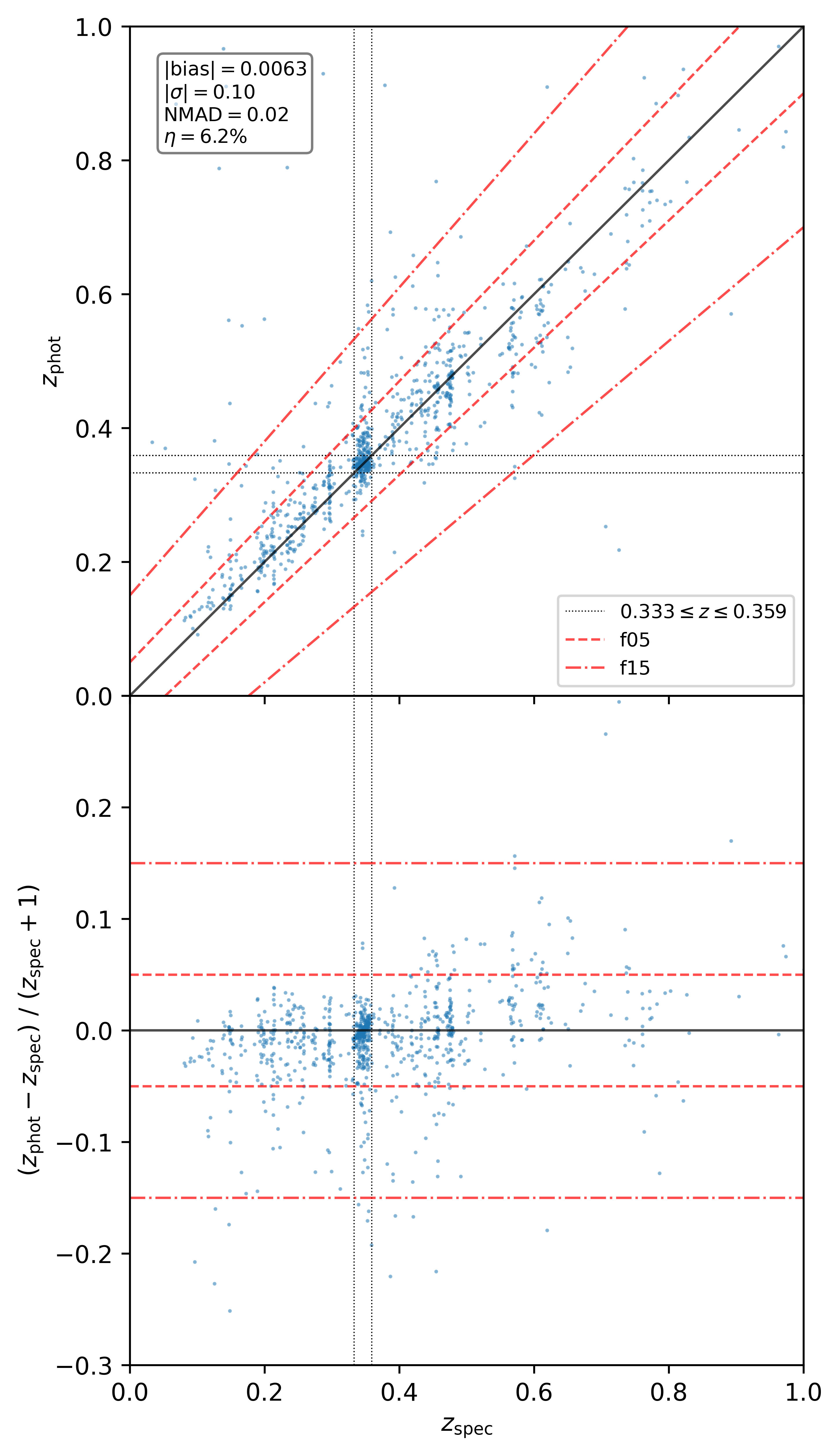}
    \caption{Spectroscopic vs photometric redshifts predicted using an RF and the "best" feature subset. The red dashed and dot-dashed lines indicate $\Delta z = $0.05 (f05) and $\Delta z = $0.15 (f15). The gray dotted lines show the spectroscopic cluster membership range.}
    \label{fig:ref_fi_zspec_vs_zphot}
\end{figure}

\end{document}